\documentclass[11pt, longbibliography, tightenlines, nofootinbib, superscriptaddress]{revtex4-2}

\usepackage[a4paper, centering, hmargin=2cm, vmargin=2.5cm]{geometry}
\usepackage{amsfonts, amsmath, amsthm, amssymb, mathrsfs, bbm, mathtools, stmaryrd}
\usepackage{array}
\usepackage{makecell}
\usepackage{xspace}
\usepackage[dvipsnames]{xcolor}
\usepackage{graphicx}
\usepackage[colorlinks=true, citecolor=cyan, linkcolor=., urlcolor=.]{hyperref}
\usepackage[nameinlink]{cleveref}
\usepackage{enumerate}
\usepackage{ragged2e}
\usepackage[font=small, labelfont=bf, width=.9\textwidth]{caption, subcaption}
\DeclareCaptionJustification{justified}{\justifying}
\usepackage{booktabs}
\usepackage{float}
\usepackage{bbold}
\usepackage{tensor}
\usepackage{physics}
\usepackage{bookmark}
\usepackage{csquotes}
\usepackage[colorinlistoftodos]{todonotes}
\usepackage{tikz}
\usepackage[compat=1.1.0]{tikz-feynman}
\usetikzlibrary{decorations.markings,calc,positioning,arrows,arrows.meta}

\newcommand{\figref}[1]{Fig.~\ref{#1}}

\NewDocumentCommand{\threedots}{mO{0}O{3pt}O{black}}{
	\begin{scope}[shift={(#1)}, rotate=#2]
		\foreach \y in {-#3, 0pt, #3} {
			\fill[#4] (\y,0) circle (0.7pt);
		}
	\end{scope}
}

\newcommand{\Pexp}{\text{$P$exp}}

\definecolor{NCBJred}{RGB}{186,20,62}
\definecolor{NCBJdarkgray}{RGB}{89,89,89}
\definecolor{NCBJlightgray}{RGB}{200,200,200}
\definecolor{duskblue}{HTML}{2c497f}
\definecolor{lavendergrey}{HTML}{808a9f}
\definecolor{grapefruitpink}{HTML}{ff6b6b}
\definecolor{crimsonviolet}{HTML}{4a001f}
\tikzset{p_node/.style={circle, fill, inner sep=1.5pt}}

\numberwithin{equation}{section}
\crefname{appendix}{appendix}{appendices}
\Crefname{appendix}{Appendix}{Appendices}
\AtBeginDocument{\addtocontents{toc}{\protect\setlength{\parskip}{0pt}}}

\begin{document}

\title{Coarse-grained models for loop quantum gravity and their renormalization}

\author{Mehdi Assanioussi}
\email{mehdi.assanioussi@ncbj.gov.pl}
\affiliation{National Centre for Nuclear Research, Department of Fundamental Research, Pasteura 7, 02-093 Warsaw, Poland}
\author{Martin Zei\ss}
\email{martin.zeiss@ncbj.gov.pl}
\affiliation{National Centre for Nuclear Research, Department of Fundamental Research, Pasteura 7, 02-093 Warsaw, Poland}

\begin{abstract}

\noindent We introduce a family of effective theories and use them to define a specific type of coarse-grained models for canonical loop quantum gravity. Each effective theory is characterized by two parameters and it mainly consists of a Hilbert space of coarse states and an effective Hamiltonian operator.
The construction of the coarse Hilbert spaces relies on a systematic coarse-graining procedure for spin network states. The effective Hamiltonians are then obtained from an analysis of the interplay between this coarse-graining procedure and the action of various Hamiltonian operators defined in loop quantum gravity. Finally, we provide a prescription for implementing the Hamiltonian non-perturbative renormalization framework for the coarse-grained models, and we explicitly write down the renormalization flow equations.
The aim of this work is to build an arena for studying the emergence of the continuum limit and the derivation of phenomenological models in the context of canonical loop quantum gravity.
	
\end{abstract}

\maketitle

\makeatletter
\let\l@subsubsection\@gobbletwo
\makeatother

\tableofcontents
\makeatletter
\let\toc@pre\relax
\let\toc@post\relax
\makeatother


\newpage


\section{Introduction}

Loop quantum gravity (LQG) \cite{Gambini:1996ik, Ashtekar:2004eh, Rovelli:2004tv, Thiemann:2007pyv} is an approach to solving the long-standing problem of developing a quantum theory that consistently describes gravity and the fields of the Standard Model. It provides a framework where gravity and matter fields are quantized in a background independent setting, with well-established mathematical foundations. Yet, the construction of the continuum limit of the theory is an open issue, and the question of how the effective smooth spacetime emerges from the loop quantum theory remains open. This step is crucial to complete the LQG program and be able to produce falsifiable predictions.

One of the most prominent ideas in the direction of resolving this issue in LQG is to develop adequate coarse-graining and renormalization schemes for the theory. These constitute a bottom-up procedure to systematically derive effective theories from the fundamental theory at the Planck scale, taking into account the quantum dynamics. There have been many works which introduce and study such methods, both in the canonical \cite{Livine:2013gna, Charles:2016xwc, Lang:2017beo, Lang:2017yxi, Lang:2017oed, Lang:2017xrb, Liegener:2020dbc, Thiemann:2020cuq, Thiemann:2022voq, Thiemann:2022ulb, Thiemann:2022ykh, Zarate:2025qlg, RodriguezZarate:2025ipb, Zarate:2025zqz} and covariant settings \cite{Dittrich:2011zh, Dittrich:2012jq, Dittrich:2014ala, Bahr:2014qza, Steinhaus:2020lgb}. 
The expectation is that efficient coarse-graining and renormalization procedures would render many of the ambiguities in the fundamental theory irrelevant to measurable observables, and they would provide accurate approximations of physical states and their dynamics. This in turn would pave the way to build predictive models that may eventually confront LQG with experiments and observations.

In the present article, we construct a family of effective theories and define a collection of coarse-grained models based on them. These effective theories are obtained using an extensive coarse-graining procedure which combines the techniques introduced in \cite{Livine:2013gna, Charles:2016xwc} with several new concepts and methods developed herein. The kinematical structure of these effective theories is characterized by the notion of \emph{multi-reflexive \mbox{$(n,p)$-complete} graphs}, and the dynamics is described through a hierarchy of \emph{interactions} among the vertices of the coarse states.
Our coarse-grained models are defined as a family of effective theories, ordered with respect to two parameters which we interpret as ultraviolet cutoffs.
Then we propose a prescription to implement the framework of Hamiltonian non-perturbative renormalization \cite{Lang:2017beo, Lang:2017yxi, Lang:2017oed, Lang:2017xrb, Liegener:2020dbc, Thiemann:2020cuq, Thiemann:2022voq, Thiemann:2022ulb, Thiemann:2022ykh, Zarate:2025qlg, RodriguezZarate:2025ipb, Zarate:2025zqz} for any coarse-grained model, and we provide the renormalization flow equations. The objective of this work is to build a rigorous setup where one could analyze the emergence of a continuum limit in canonical LQG through the study of renormalization flows.

The article is organized as follows. Section II briefly reviews the framework of canonical loop quantum gravity. Section \ref{sec:coarse-graining} develops the coarse-graining procedure, leading from the kinematical Hilbert space of LQG to the family of coarse Hilbert spaces $\mathcal H_{(n,p)}$, labeled by two resolution parameters $n$ and $p$. This procedure consists of four steps: (i) passing from states with embedded graphs to states with abstract graphs, with or without geometric tags that we define here; (ii) coarse-graining vertices via gauge fixing, following \cite{Charles:2016xwc}; (iii) coarse-graining the multiple edges connecting any pair of coarse vertices via a method presented here; and (iv) coarse-graining the newly introduced geometric tags. Combining these four steps, we construct the coarse-graining maps $\mathcal I_{n,p}$ and the resulting kinematical structures.
Section \ref{sec:Effective Dynamics} addresses the effective dynamics on these coarse Hilbert spaces. We examine how the fundamental Hamiltonian operators of canonical LQG act under the coarse-graining procedure, and use this analysis to identify and classify the resulting effective interactions among the vertices of a coarse state. Based on this classification, we propose an effective Hamiltonian on each coarse space $\mathcal H_{(n,p)}$, given as a sum over all admissible interactions with independent coupling coefficients; these coefficients will later serve as the parameters of the renormalization flow. The section concludes with our definition of effective theories.
Section \ref{sec:Hamiltonian_Renormalization} introduces the coarse-grained models and presents a prescription to implement the Hamiltonian non-perturbative renormalization framework for our models. We construct a family of refinement maps between the coarse Hilbert spaces, and use them to write down the corresponding renormalization flow equations which are to be solved for the coupling coefficients of the effective Hamiltonian. We close the article with a summary of these results and a discussion of possible directions for future work.



\section{Loop quantum gravity in a nutshell}\label{sec:LQG}

Loop quantum gravity \cite{Gambini:1996ik, Ashtekar:2004eh, Rovelli:2004tv, Thiemann:2007pyv} is a background independent approach to quantizing general relativity based on the Ashtekar-Barbero formulation. 
In this canonical formulation, general relativity is formulated as a gauge theory, where the phase space variables are a $su(2)$-algebra-valued connection $A_a^i$, called the Ashtekar-Barbero connection, and the densitized triad $E_i^a$, $a$ being a space index and $i$ an algebra index.
This formulation comes with three first-class constraints: the Gauss constraint imposing $SU(2)$ gauge invariance, the vector constraint imposing spatial diffeomorphism invariance on the spatial hypersurface, and finally the scalar or Hamiltonian constraint imposing invariance under diffeomorphisms orthogonal to the hypersurface.
The requirement of background independence prevents quantizing the variables $(A,E)$ directly, since they satisfy a singular Poisson algebra which involves a Dirac delta distribution.
One proceeds instead by constructing an algebra of smeared variables that still parametrize the same phase space. 
Inspired by lattice gauge theory, loop quantum gravity proceeds by quantizing the holonomy-flux algebra obtained from the Ashtekar-Barbero variables.

Given an oriented $1$-dimensional path $e$ (edge) and a $2$-dimensional surface $S$ embedded in the spatial hypersurface $\Sigma$, the holonomy $h_e[A]$ and the flux $P_{S,\xi}$, associated to the edge $e$ and the surface $S$ respectively, are defined as
\begin{align}
	h_e[A] = \Pexp \left(-\int_e dx^a A^i_a\tau_i \right) \ , \qquad
	P_{S,\xi} = \frac{1}{2}\int_S dx^b\wedge dx^c\,\epsilon_{abc}\xi^i(x) {E}^a_i(x)\ ,
\end{align}
where $\{\tau_i\}$ is an orthonormal basis of the $su(2)$ algebra, $\Pexp$ denotes the path-ordered exponential and $\xi$ is a smooth smearing function valued in $su(2)$.
The advantage of using holonomies and fluxes is that they transform in a closed and simple fashion under the action of $SU(2)$ gauge transformations and spatial diffeomorphisms. Namely, for any $g\in SU(2)$ we have
\begin{align}
	h_e[A] \rightarrow g_{t(e)}h_e[A]g^{-1}_{s(e)} \ ,\qquad
	P_{S,\xi} \rightarrow P_{S,g^{-1} \xi g} \ ,
\end{align}
where $s(e)$ and $t(e)$ denote the starting (source) point and end (target) point of the oriented path $e$, respectively; and the action of a spatial diffeomorphism $\phi$ gives
\begin{align}
	h_e[A] \rightarrow h_{\phi(e)}[A] \ ,\qquad
	P_{S,\xi} \rightarrow P_{\phi(S), \phi^{-1}(\xi)}\ .
	\label{GaugeTransf}
\end{align}

The quantization of the holonomy-flux algebra provides a vector space $\rm{Cyl}$ of cylindrical functions of the connection variable $A$.
These are complex-valued functions which depend on the connection through finitely many holonomies:
\begin{equation}
	\Psi[A] \equiv \psi(h_{e_1}[A],\ldots ,h_{e_w}[A]) \ ,
\end{equation} 
with $\psi : {\rm SU}(2)^w \rightarrow \mathbbm{C}$. The set of embedded edges $\Gamma \equiv \{e_1,...,e_w\}$ is called the graph of $\Psi$.
The space of all cylindrical functions with a graph $\Gamma$ is denoted ${{\rm Cyl}}_\Gamma$ and we have ${\rm Cyl} \equiv  \cup_\Gamma\ {\rm Cyl}_\Gamma$.
The kinematical Hilbert space ${\mathcal H}_{\rm kin}$ of LQG is then defined as the completion of ${\rm Cyl}$ with respect to the inner product induced by the Haar measure $\mu_H$ on $SU(2)$ \cite{Ashtekar:1993wf, Ashtekar:1994mh, Marolf:1994cj, Ashtekar:1995zh, Lewandowski:2005jk}, with a vacuum state $\Omega_{\rm AL}$ called the \emph{Ashtekar-Lewandowski vacuum}.
Furthermore, the Hilbert space of a single edge $e$ corresponds to $L^2(SU(2),d\mu_H)$. Using the Peter-Weyl theorem, an orthonormal basis for $L^2(SU(2),d\mu_H)$ is given by normalized matrix elements of the irreducible representations $D^{(j)}(h_e)$ of $SU(2)$, labeled by spins $j$.
It follows that the Hilbert space ${\mathcal H}_{\rm kin}$ admits an orthonormal basis consisting of spin network functions: such a function is defined by an embedded graph $\Gamma$, with spins (labeling $SU(2)$ irreducible representations) assigned to the edges $\{e_i\}$ and $SU(2)$ tensors (which couple the spins) assigned to the vertices $\{v_i\}$.

The Hilbert space ${\mathcal H}_{\rm kin}$ is then expressed as the direct sum
\begin{align}\label{decomp} 
{\mathcal H}_{\rm kin} = \bigoplus_{\Gamma} {\mathcal H}_\Gamma \ ,
\end{align}
where $\Gamma$ ranges over all the classes of embedded graphs \cite{Ashtekar:2004eh} and ${\mathcal H}_\Gamma$ is the Hilbert space defined as the completion of the space ${\rm Cyl}_\Gamma$, where no spin label is equal to zero. Each of these Hilbert spaces ${\mathcal H}_\Gamma$ is of the form $L^2 (SU(2)^{w_e}, d\mu_H)$, where $w_e$ is the number of edges of $\Gamma$.

As for the operator algebra acting on ${\mathcal H}_{\rm kin}$, every cylindrical function $\Psi$ (including single holonomies) defines a multiplication operator, while the quantum flux acts as a derivative operator
\begin{align}
	\hat{P}_{S,\xi}\  =\ \frac{\kappa\beta\hbar}{2}\sum_{x\in S}\xi^i(x)\sum_{e} \omega_S(e)\hat{J}_{x,e,i}\ ,
\end{align}
where $\kappa = 8\pi G$, $G$ being Newton's gravitational constant and $\beta$ is the Immirzi-Barbero parameter. The sum over $e$ runs through the edges at $x$ and $\omega_S(e) \in \{-1,0,1\}$, depending on whether $e$ goes down, along, or up respectively, with respect to the surface $S$.
The operator $\hat{J}_{x,e,i}$ is assigned to a pair $(x,e)$ and corresponds to either the $SU(2)$ left- or right-invariant vector field, depending on whether the edge $e$ begins or ends at the point $x$ respectively.

The Gauss and spatial diffeomorphism constraints in loop quantum gravity can be implemented through \emph{group averaging techniques} \cite{Ashtekar:1995zh}. The central idea is to construct gauge invariant states by averaging kinematical states over the action of the symmetry group generated by the constraints. We first recall the general averaging procedure. Given a state $\Psi$ and a group $G$ with unitary action $U(g)$, the averaging map takes the form
\begin{align}
\int_G dg \, U(g)\Psi \ ,
\end{align}
where $dg$ is an invariant measure on $G$. In cases where no suitable measure exists, one attempts to construct a rigging map with definite action on the kinematical states. This is for instance the case for spatial diffeomorphisms. Either way, the resulting states are invariant under the group action and thus form the space of solutions to the quantum constraint.

For the Gauss constraint, the symmetry group is the local $SU(2)$ gauge group. Averaging over gauge transformations projects the kinematical Hilbert space $\mathcal{H}_{\rm kin}$ onto the subspace of gauge invariant states, yielding the Hilbert space $\mathcal{H}_{\rm G} \subset \mathcal{H}_{\rm kin}$. Its basis is provided by $SU(2)$ gauge invariant spin network states, with invariant tensors at the vertices called intertwiners.

The spatial diffeomorphism invariance is implemented analogously by averaging over the group of spatial diffeomorphisms $\text{Diff}(\Sigma)$ acting on the hypersurface $\Sigma$, but with some alterations. Indeed, since the spatial diffeomorphism group is infinite-dimensional and non-compact, no natural invariant measure is available, and the averaging must therefore be defined through a rigging map with an image in the algebraic dual $\rm{Cyl}^*$ of the space of cylindrical functions $\rm{Cyl}$. Denoting $\text{Diff}_\Gamma(\Sigma)$ the set of diffeomorphisms in $\text{Diff}(\Sigma)$ which preserve the graph $\Gamma$, one defines the rigging map $\eta$ as
\begin{align}
\eta(\Psi)=\sum_{\phi \in \text{Diff}(\Sigma)/\text{Diff}_\Gamma(\Sigma)}
[U_\phi \Psi]^* ,
\label{rigging.map}
\end{align}
which associates to a cylindrical function in $\rm{Cyl}$ a diffeomorphism invariant distribution in $\rm{Cyl}^*$. The rigging map is also used to define an inner product on the resulting space of states, given by the Ashtekar-Lewandowski measure \cite{Ashtekar:1993wf, Ashtekar:1994mh, Marolf:1994cj, Ashtekar:1995zh, Lewandowski:2005jk}, leading to the Hilbert space
\begin{align}
\mathcal{H}_{\rm diff} \subset \rm{Cyl}^*
\end{align}
of states invariant under both $SU(2)$ gauge transformations and spatial diffeomorphisms. In this way, group averaging provides a systematic implementation of first-class constraints that generate a symmetry group.

Unlike the other two constraints discussed above, the Hamiltonian constraint does not generate a group and thus group averaging techniques cannot be applied to construct the space of solutions. The alternative avenue is to build a quantum operator, densely defined on one of the Hilbert spaces presented above, then use it to define a constraint equation for the physical states. This is the approach in the fully constrained theory. However, with all the proposals for the Hamiltonian operator present in the literature \cite{Thiemann:1996aw, Thiemann:2003zv, Giesel:2006uj, Alesci:2015wla, Assanioussi:2015gka}, one is unable to completely solve the constraint equation and to arrive at the full physical Hilbert space. Non-trivial solutions can be derived, but the non-polynomial form and the complexity of the action of the Hamiltonian operators prevent the construction of a full space of solutions. This motivated an alternative approach to the problem of evolution in gravity: using matter fields as reference frames. This is the deparametrization of gravity \cite{Kuchar:1990vy, Rovelli:1993bm, Brown:1994py, Kuchar:1995xn, Giesel:2012rb, Giesel:2016gxq}, which comes at the cost of choosing a specific reference frame to parametrize either time or both time and space. Hence the description and interpretation of the physics derived within a given model are tied to the choice of frame. Nevertheless, deparametrization allows one to bypass the issue of the quantum Hamiltonian constraint equation, and in many cases, it leads to complete quantum models where gravity is fully quantized \cite{Giesel:2012rb, Giesel:2016gxq, Assanioussi:2017tql}. As such, these models become a very fertile ground to develop new methods and ideas to answer even more complex questions concerning the semi-classical and continuum limit in loop quantum gravity.
In section \ref{sec:Effective Dynamics} of the present article, we propose an effective dynamics for the coarse states, under the assumption that we are in a deparametrized setting where there is a chosen time parameter. Therefore, when we discuss the Hamiltonian operators, these should be understood as true Hamiltonians generating the evolution of the physical states and the observables with respect to a given time.

It is important to emphasize that the loop quantum gravity program provides a framework where the fields of the Standard Model can also be quantized in a background independent fashion \cite{Gambini:1996ik, Thiemann:2007pyv}. In fact, the methods described above can be applied to any gauge theory with a compact gauge group; in particular, the quantization of gravity coupled to Yang-Mills fields is straightforward. In this case, the structure of the states remains similar to that of vacuum gravity, namely, basis states are labeled by graphs (up to diffeomorphisms), irreducible representations of the gauge groups at the edges, and gauge invariant intertwiners at the vertices. Scalar and fermionic fields are included via some modifications of the kinematical structures, but the quantization program can be completed (see \cite{Ashtekar:2002vh, Kaminski:2005nc, Kaminski:2006ta, Han:2006iqa, Domagala:2010bm, Domagala:2012tq} and \cite{Morales-Tecotl:1995oxb, Baez:1997bw, Thiemann:1997rq, Thiemann:1997rt, Mansuroglu:2020acg, Lewandowski:2021bkt}). Along the same lines, we stress that the coarse-graining methods and constructions described in this article also apply to loop quantum models where gravity is coupled to other gauge fields, in particular Yang-Mills fields. The extensions to the cases where scalar fields and fermions are present will be investigated in the future.

As mentioned in the introduction, the construction of the effective theories that we propose relies on a coarse-graining procedure which consists of two parts: first, we coarse-grain kinematical states to obtain coarse Hilbert spaces, then we identify the mappings induced by coarse-graining the action of the fundamental Hamiltonian operators in order to define an effective dynamics.
In the following section, we develop in detail the coarse-graining procedure for the kinematical states which consists of four steps. First, we discuss the passage from states with embedded graphs to states with abstract graphs, with the possibility of retaining selected geometric information, that is, the \emph{geometric tags} (sec.~\ref{sec:abstract graphs}). Second, we review the coarse-graining procedure for vertices via gauge fixing \cite{Charles:2016xwc} to arrive at the \emph{loopy spin networks} (sec.~\ref{sec:coarse-graining vertices}). We then introduce a procedure to coarse-grain multiple edges connecting a pair of vertices. Next, we provide a prescription to coarse-grain the geometric tags. Finally, we define our coarse-graining maps and identify the kinematical structures of the effective theories that form the basis of our coarse-grained models.


\section{Coarse-graining spin network states}\label{sec:coarse-graining}


\subsection{Abstract graphs with and without geometric tags}\label{sec:abstract graphs}

As presented in the previous section, the kinematical structure of canonical LQG consists of the Hilbert space of gauge invariant states $\mathcal{H}_{\rm G}$, or the space of gauge and spatial diffeomorphism invariant states $\mathcal{H}_{\rm diff}$. The Hilbert space of states with abstract graphs does not depend on which Hilbert space we start from; both $\mathcal{H}_{\rm G}$ and $\mathcal{H}_{\rm diff}$ are suitable. All the discussions and reasoning that follow can be applied to both spaces. We are interested in abstract graphs because their combinatorial structure is much more suitable for the process of coarse-graining. Furthermore, the restriction to combinatorial structures provides the advantage that, unlike $\mathcal{H}_{\rm G}$ and $\mathcal{H}_{\rm diff}$, the Hilbert space of states with abstract graphs is separable.
Indeed, the Hilbert space $\mathcal{H}_{\rm G}$ admits the set of spin network states as a basis; however, this basis is not countable because the set of embedded graphs is not countable. Thus the space $\mathcal{H}_{\rm G}$ is not separable. Averaging with respect to spatial (semi-analytic or smooth) diffeomorphisms does not change the situation, and $\mathcal{H}_{\rm diff}$ is also not separable \cite{Baez:1993wg, Grot:1996kj}.

The non-separability of the Hilbert space has been viewed as a significant disadvantage. It often stands in the way of applying many standard physical and mathematical methods of quantum theory, which could help in understanding the physics that the theory describes. Several proposals exist for constructing a separable Hilbert space for LQG. One idea is to average with respect to piecewise analytic homeomorphisms \cite{Zapata:1997db, Zapata:1997da}. Another idea is based on the observation that the space $\mathcal{H}_{\rm diff}$ consists of an uncountable sum over the moduli of mutually isomorphic separable Hilbert spaces \cite{Fairbairn:2004qe}. As it turns out, each of these separable Hilbert spaces is preserved by all diffeomorphism invariant operators constructed in the theory so far. This means that any algebra of diffeomorphism invariant observables would select only one of these spaces and it is irrelevant which one it is. Hence, the proposal is to simply select one of these spaces as the diffeomorphism invariant Hilbert space of the theory. The appeal of working with states defined with abstract graphs was one of the motivations for introducing the \emph{algebraic quantum gravity} program, in which many interesting semi-classical results with coherent-state methods have been derived \cite{Giesel:2006uj, Giesel:2006uk, Giesel:2006um, Giesel:2007wn, Dapor:2017rwv, Dapor:2017gdk}.  

In the interest of building effective models, working with a separable Hilbert space is certainly an advantage, especially if we are to consider the deparametrized theory, where one chooses a frame from the start.
Our approach to arriving at separable Hilbert spaces is to modify the structure of the graphs associated to the states, by moving from embedded graphs, with uncountable geometric information, to abstract graphs without any, or with very limited, geometric information, as we explain below. Our interpretation of this procedure is that we coarse-grain the geometric information encoded in embedded graphs, and retain only the minimal amount needed to represent the algebra of observables of interest on a separable Hilbert space. In canonical LQG, there are several geometric observables, with concrete physical interpretation, which depend on the differential structure of the graph, in particular the tangent vectors of the edges at the vertices of the graph. Indeed, observables such as (one of the versions of) the volume \cite{Ashtekar:1997fb}, Ricci curvature \cite{Alesci:2014aza} and the Hamiltonian operators include some geometric information in their definitions.
For our construction, an abstract graph consists of a finite set of vertices and a finite set of oriented edges connecting these vertices, including self-loops (reflexive edges). The vertices of an abstract graph are at least three-valent, except for independent single self-loops, in which case two-valent vertices are present. The set of such abstract graphs is therefore countable.

Given an abstract graph, one can build a space of cylindrical functions associated to it, similarly to the construction of the standard ${{\rm Cyl}}_\Gamma$ described earlier. Following the same steps to obtain $\mathcal{H}_{\rm G}$, one arrives at a Hilbert space $\mathcal{H}_{\rm A}$ of cylindrical functions associated to abstract graphs. This space admits a basis formed of abstract gauge invariant spin network functions, almost identical to the original gauge invariant spin network states, with the difference that the embedded graphs are replaced by abstract ones. Since the set of abstract graphs is countable, the Hilbert space $\mathcal{H}_{\rm A}$ is separable. 
Another way to arrive at $\mathcal{H}_{\rm A}$ directly from $\mathcal{H}_{\rm G}$ is to perform an averaging of the states with respect to the group of homeomorphisms which preserve only the combinatorial data of the graph, namely the set of vertices and their connectivity through the edges, without modifying the algebraic data, i.e.~spins and intertwiners. This would be achieved via a rigging map $\zeta$, analogous to the one used to construct the space of diffeomorphism invariant states $\mathcal{H}_{\rm diff}$ in \eqref{rigging.map}, acting on the span $\rm{Cyl}_G$ of spin network states in $\mathcal{H}_{\rm G}$.
Note that the space $\mathcal{H}_{\rm A}$ corresponds to the boundary Hilbert space in the \emph{spin foam} models approach \cite{Perez:2012wv, Rovelli:2014ssa, Engle:2023qsu}, which attempts to define a covariant formulation of loop quantum gravity. The structure of the states in $\mathcal{H}_{\rm A}$ is also the basis of the \emph{group field theory} formalism \cite{Oriti:2006se, Oriti:2013aqa}, where the states are organized in a Fock space setup.

We can go a step further and instead of having purely algebraic and combinatorial data, additionally retain some data of geometric nature. As mentioned earlier, several observables depend on some amount of geometric information provided by the embedded graphs. Hence, at this stage, if one does not want to alter significantly the original observables and their algebra, it would be reasonable to keep a minimal amount of geometric information. The main restriction on such information is that it does not spoil the separability of the Hilbert space. In order to guarantee this, the geometric data to be retained must form a countable set, such that each basis state is characterized by a finite set of such geometric data. 
Looking at the expressions of the observables available in the LQG literature, one observes that the only geometric information that is used in the definitions of the operators consist of two elements. The first is whether two edges at a vertex are tangent or not; the second is the orientation of an ordered triple of edges at a vertex, which corresponds to the sign of the determinant of the matrix given by the ordered tangent vectors of the three edges. Having an embedded graph with a vertex $v$ and a set of edges $\mathfrak{e}_v=\{e_i\}$ meeting at $v$ with tangent vectors $\{\dot e_i \in \mathbbm{R}^3\}$, one can define two maps $\vartheta_v$ and $\varepsilon_v$ as
\begin{alignat}{2}\label{Geo_Tags}
	\vartheta_v: \mathfrak{e}_v \times \mathfrak{e}_v &\longrightarrow \{0,1\} \qquad &,\qquad \varepsilon_v: \mathfrak{e}_v \times \mathfrak{e}_v \times \mathfrak{e}_v &\longrightarrow \{-1,0,1\} \\[1em]
	\vartheta_v(e_i, e_j) &= 
	\begin{cases}
		0\ ,\ \text{if $\dot e_i \wedge \dot e_j = 0$} \\ 1\ ,\ \text{otherwise}
	\end{cases}
	&\varepsilon_v(e_i, e_j, e_k) &= \text{sgn}(\text{det}[\dot e_i, \dot e_j, \dot e_k]) \in \{-1,0,1\} \ . \notag
\end{alignat}
These two maps encode all the geometric information required by the gauge and diffeomorphism invariant observables considered in LQG so far. For each vertex of a graph, the maps $\vartheta_v$ and $\varepsilon_v$ would assign an additional finite set of data encoding minimal geometric information, which we call a \emph{geometric tag}. At each vertex $v$, this geometric tag would consist of two sets of discrete values, one associated to the pairs of edges at the vertex while the second is associated to triples of edges.
Since the maps $\vartheta_v$ and $\varepsilon_v$ have finite sets as images, the set of geometric tags is countable. Thus, the addition of the geometric tags to the abstract graph structure does not ruin the countability of the set of graphs. Consequently, the Hilbert space $\mathcal{H}_{\rm T}$ spanned by spin network states with \emph{tagged abstract graphs} is separable. Analogously to $\mathcal{H}_{\rm A}$, one can build this Hilbert space by first defining homeomorphisms which preserve both the combinatorial data and the data provided by the maps $\vartheta_v$ and $\varepsilon_v$ at each vertex of the graph; then introducing a rigging map to perform an averaging of the states with respect to this group of homeomorphisms.

For all the results and constructions developed in the remainder of the present article, we assume starting from the space $\mathcal{H}_{\rm T}$ and we consider states with geometric tags. This is the more general case, and the restriction to the space $\mathcal{H}_{\rm A}$ is straightforward as one simply drops the geometric elements in all the structures we introduce.

In the following, we describe how to coarse-grain the vertices of states in $\mathcal{H}_{\rm T}$ via the method of gauge fixing, leading to the so called loopy spin networks introduced in \cite{Charles:2016xwc}.


\subsection{Coarse-graining vertices: revisiting the loopy spin networks}\label{sec:coarse-graining vertices}

We begin by reviewing the method of coarse-graining the vertices of a graph via gauge-fixing introduced in \cite{Charles:2016xwc} and \cite{Livine:2013gna} (which are based on \cite{Freidel:2002xb}). This is the second step in the coarse-graining procedure and it provides the basic concepts and techniques for coarse-graining states in the Hilbert space $\mathcal H_T$.

The core idea of coarse-graining states via gauge-fixing is to be able to encode the curvature of the Ashtekar-Barbero connection, carried by the loops present in some bounded region of a given graph, into a single vertex with self-loops. The self-loops account for the gravitational excitations within the initial bounded region. Given a gauge invariant cylindrical function $\Psi$ with graph $\Gamma$, the procedure relies on using $SU(2)$ gauge transformations to trivialize a set of holonomies associated to a subgraph of $\Gamma$. Namely, using equation \eqref{GaugeTransf} for the transformation of a holonomy under $SU(2)$ gauge transformations, gauge invariance of $\Psi$ implies
\begin{equation}
	\Psi(\{\dots, g_{t(e)} h_e g^{-1}_{s(e)} ,\dots \}) = \Psi(\{\dots, h_e , \dots \})\ .
\end{equation}
In particular, if one chooses $g_{s(e)} = \mathbb 1$ and $g_{t(e)} = h_e^{-1}$, the holonomy associated to the edge $e$ becomes the identity element on $SU(2)$. This is the trivialization of the holonomy $h_e$.

\begin{figure}[h]
	\centering
	\begin{tikzpicture}[
		scale=0.7,
		v/.style={circle,fill=black,inner sep=1.5pt},
		rv/.style={circle,fill=NCBJred,inner sep=1.5pt},
		gv/.style={circle,fill=NCBJdarkgray,draw,inner sep=1.5pt},
		sv/.style={circle,fill=NCBJlightgray,draw,inner sep=1.5pt},
		every loop/.style={very thick,blue!80!black,looseness=25},
		dashedcircle/.style={draw,NCBJdarkgray,dashed,very thick},
		>={Latex[length=2.5mm]}
		]
		
		
		\node[v] (A) {};
		\node[v] (B) at ($(A)+(1.6,0)$) {};
		\node[rv] (C) at ($(B)+(2.2,0)$) {};
		\node[v] (D) at ($(C)+(1.6,0)$) {};
		
		\node (Gamma) at ($(B)+(0,1)$) {$\gamma_i$};
		
		\node[rv] (E) at ($(A)+(0,-1.4)$) {};
		\node[v] (F) at ($(B)+(0,-1.4)$) {};
		\node[v] (G) at ($(C)+(0,-1.4)$) {};
		
		\node[rv] (J) at ($(F)+(0.7,-2.0)$) {};
		
		\draw[very thick] (B)--(C);
		\draw[very thick] (F)--(G);
		\draw[very thick] (G)--(J);
		
		\draw[very thick] (B) to[bend left] (J); 
		\draw[very thick] (F) -- (J); 
		\draw[very thick] (E) to[bend right] (J); 
		
		\draw[NCBJred, very thick] (A)--(E)--(F);
		\draw[NCBJred, very thick] (E)--(B);
		\draw[blue!80!black, very thick] (A)--(B);
		\draw[blue!80!black, very thick] (B)--(F);
		
		\draw[NCBJred, very thick] (G)--(C)--(D);
		\draw[blue!80!black, very thick] (G) to[out=60,in=-150] (D);
		\draw[blue!80!black, very thick] (G) to[out=30,in=-120] (D);
		
		\draw[very thick] (A) -- ++(-0.8,0.8);
		\draw[very thick] (D) -- ++(0.8,0.8);
		\draw[very thick] (J) -- ++(0,-1);
		
		\draw[dashedcircle] ($(A)!0.5!(F)$) circle (1.3);
		\draw[dashedcircle] ($(D)!0.5!(G)$) circle (1.3);
		\draw[dashedcircle] ($(J)$) circle (0.8);
		
		
		\draw[->,very thick] ($(D)+(1.6,-2)$) -- ++(2.2,0);
		
		
		\begin{scope}[shift={($(D)+(6,-2)$)}]
			\node[rv] (R1) at (0,1.3) {};
			\node[rv] (R2) at ($(R1)+(3.6,0)$) {};
			\node[rv] (R4) at ($(R1)+(1.8,-2.6)$) {};
			
			\draw[very thick] (R1) -- ++(-0.9,0.9);
			\draw[very thick] (R2) -- ++(0.9,0.9);
			\draw[very thick] (R4) -- ++(0,-1);
			
			\draw[very thick, blue!80!black] (R1) edge[out=-95,in=-140,looseness=50] (R1);
			\draw[very thick, blue!80!black] (R1) edge[out=160,in=-155,looseness=50] (R1);
			\draw[very thick, blue!80!black] (R2) edge[out=20,in=-25,looseness=50] (R2);
			\draw[very thick, blue!80!black] (R2) edge[out=-40,in=-85,looseness=50] (R2);
			
			\draw[very thick] (R1) to[bend left=25] (R2);
			\draw[very thick] (R1) to[bend right=25] (R2);
			
			\draw[very thick] (R1) to[bend left=25] (R4);
			\draw[very thick] (R1) -- (R4);
			\draw[very thick] (R1) to[bend right=25] (R4);
			
			\draw[very thick] (R2) -- (R4);
		\end{scope}
	\end{tikzpicture}
	\vspace*{5mm}
	\caption{Coarse-graining vertices via gauge fixing: On the left-side, the subgraphs $\gamma_i$ are encircled by gray dashed lines, the associated root vertices and trees are in red, the internal edges of the subgraphs are in blue, while the external edges are in black. On the right side is the coarse-grained graph where the edges of the trees have been removed, and the internal edges have become self-loops.}
	\label{fig:CG_via_GF}
\end{figure}
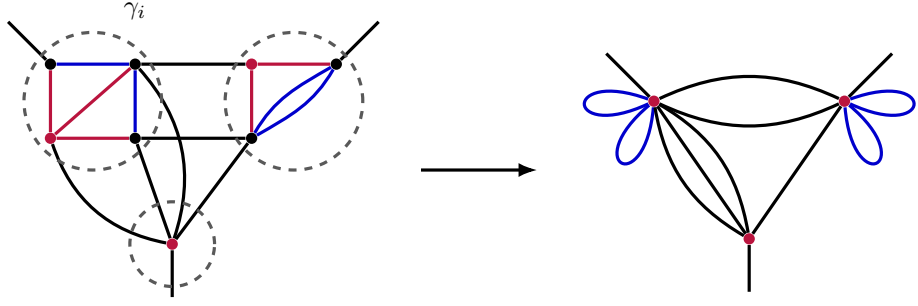

We now provide the prescription to perform the coarse-graining of vertices, described in detail in \cite{Charles:2016xwc}. Given an arbitrary spin network state with tagged abstract graph $\Gamma$, the procedure below maps it to a \emph{coarse} spin network state with a fixed number of vertices $n>0$. If the initial state has a number of vertices $\nu_\Gamma$ less than or equal to $n$, then it is unaffected by the coarse-graining process and it is simply mapped to itself. This procedure, illustrated in \figref{fig:CG_via_GF}, goes as follows:
\begin{itemize}
	\itemsep0em 
	
	\item[i.] select a finite number $n \in \mathbbm N^+$ of disjoint subgraphs $\{\gamma_i\}$, such that the union $\bigcup_{i=1}^n \gamma_i$ contains all the vertices of $\Gamma$;
	
	\item[ii.] in each subgraph $\gamma_i$, select a vertex $v_i$, called the root vertex, and a maximal tree $T_i$ starting at $v_i$. A maximal tree $T$ of a graph $\gamma$ is a connected subgraph containing all the vertices of $\gamma$, with no self-loops, such that there is a unique path in $T$ connecting any pair of vertices in $\gamma$;
	
	\item[iii.] for each subgraph $\gamma_i$, use gauge transformations at the vertices as mentioned above in order to gauge fix all the holonomies associated to the edges of the tree $T_i$ to the identity $\mathbb 1$. This can be achieved by performing gauge transformations iteratively through the vertices of the tree, starting from the root vertex $v_i$;
	
	\item[iv.] collapse each subgraph $\gamma_i$ to its associated root vertex $v_i$: all the edges of the tree $T_i$ in $\gamma_i$ are removed and the remaining ones, which we call \emph{internal edges}, become self-loops at $v_i$. The edges connecting the subgraphs $\gamma_i$ now become edges connecting the \emph{coarse vertices} $v_i$, and we call them \emph{external edges}.
\end{itemize}
Note that in the case where the graph within a chosen region consists of disconnected components, the coarse-graining is performed by collapsing the disconnected components along trivial edges, i.e.~edges with holonomies in the trivial representation (spin zero). More systematically, to build a maximal tree in such a region, one begins by choosing a maximal tree in each component, then completes the union of these trees using trivial edges connecting one tree to another, without creating loops, to arrive at the final tree. The rest of the procedure follows the steps described above.

The first step of the above procedure involves selecting the regions of the spin network graph to be coarse-grained to single vertices. In this procedure, the number $n$ of these regions and the associated subgraphs is an input that one has to choose. From a physical perspective, the number $n$ of selected regions could be interpreted as reflecting the finite resolution that a given observer has when probing the quantum geometry, such that a larger number of regions means higher resolution. 
The second and third steps are the key processes which reduce the information representing the field excitations within a given region. These steps use the available gauge freedom in order to trivialize a subset of the holonomies defining the state. The choice of root vertex and maximal tree is not unique, and at this stage the image of a spin network state via this procedure depends on these choices. We will come back to this point in section \ref{sec:CG maps and spaces}, where we propose a way to remove these dependencies via an averaging over the choices, which in turn allows us to define the coarse-graining maps. The last step concludes the procedure by providing a coarse state consisting of the root vertices with self-loops, which we now call \emph{coarse vertices}. These self-loops represent the independent loops existing in each of the original subgraphs $\gamma_i$, and they are what retains the curvature present in these regions. In fact, this coarse-graining procedure preserves the number of independent loops.

This last step also defines the intertwiner structure at each coarse vertex. For instance, given two vertices $v$ and $\tilde v$ connected by an edge $e$ and having two intertwiners $\iota_v$ and $\iota_{\tilde v}$ respectively, if one coarse-grains the two vertices to the vertex $v$ by trivializing the holonomy $h_e$, i.e.~the tree consists of the edge $e$, then the two intertwiners are mapped as
\begin{equation}\label{eq:vertex intertwiner gf}
    (\iota_v)\indices{_{\ldots \mu_e\ldots}}[h_e]\indices{^{\mu_e}_{\nu_e}}(\iota_{\tilde{v}})^{\ldots \nu_e\ldots} \quad\xrightarrow{\qquad}\quad (\hat \iota_v)\indices{^{\ldots}_{\ldots}}=(\iota_v)\indices{_{\ldots \mu_e\ldots}}\delta^{\mu_e}_{\nu_e}(\iota_{\tilde{v}})^{\ldots \nu_e\ldots}\ ,
\end{equation}
where $\hat \iota_v$ is the new intertwiner at the vertex $v$. This can be applied to any tree with an arbitrary number of vertices, and the intertwiner at the coarse vertex after collapsing the subgraph would be obtained by contracting the intertwiners of all the vertices with identity matrices (Kronecker deltas) replacing the holonomies associated to the tree. As for the internal and external edges, the holonomies associated to them are also transformed by the gauge transformation performed to trivialize the holonomies of the tree, but this is exactly what enables preserving the information about the curvature encoded in the loops of the initial graph.

By definition, coarse-graining implies that we do not intend to retain all the information encoded in the initial states. In the context of the procedure described above, the authors of \cite{Charles:2016xwc} proposed several options regarding the information removed by coarse-graining, leading to a hierarchy of states with different characteristics. In the present article, we use the loopy spin network states, which are the coarse states obtained once the information about the maximal trees is dropped. The structure of the loopy spin networks is in fact the same as that of the standard abstract spin network states, with arbitrary numbers of self-loops attached to the vertices. The Hilbert space of the loopy spin networks has been thoroughly studied and more details can be found in \cite{Charles:2016xwc}.

As noted in \cite{Charles:2016xwc} and depicted in \figref{fig:CG_via_GF}, this procedure generically leads to multiple edges connecting each pair of coarse vertices. Therefore, we generally end up with a graph structure where the external edges form loops between the coarse vertices. These loops carry other curvature contributions which have not been fully coarse-grained. The authors of \cite{Charles:2016xwc} put forward a proposal on how to coarse-grain these loops in the context of loopy spin networks: the external edges between two vertices are coarse-grained to two edges connecting the same two vertices to a third new vertex with self-loops. These self-loops account for the curvature of the initial loops formed by the external edges. In the present work, we are also interested in coarse-graining the loops formed by external edges. However, we propose a different approach that, on the one hand, does not generate additional coarse vertices and, on the other hand, anticipates the implementation of dynamics for the coarse states. We discuss these aspects in more detail in section \ref{sec:Effective Dynamics}. In the following, we present our proposal for coarse-graining multiple external edges connecting two coarse vertices in the graph of a spin network state.


\subsection{Coarse-graining edges of spin networks}\label{sec:coarse-graining edges}

Before presenting the details, we stress that the objective of the procedure is to systematically reduce the number of edges connecting each pair of coarse vertices, while preserving the number of independent loops formed by these edges and the curvature they encode. The technical aspect relies on the same idea of gauge fixing used to coarse-grain the vertices.

The procedure maps an arbitrary number of edges connecting two vertices of a graph to some chosen fixed number $p$ of \emph{coarse edges}, with additional self-loops at these vertices. The cases where we start with at most $p$ edges connecting a pair of vertices are trivial, in the sense that no coarse-graining of these edges is necessary. Therefore, the procedure which we present here is only applied when we have more than $p$ edges connecting two vertices.

While the procedure we present below can be applied for any $p \in \mathbbm N^+$, one might argue that it is preferable to consider only the range $p \geq 2$. As we mentioned earlier, this is in anticipation of the dynamical interactions that we would like to implement in our models. Namely, the presence of more than one edge between a pair of vertices allows defining graph-preserving dynamics involving only these two vertices. This interaction would be defined through the holonomies around the loops formed by these edges. Hence, our intuition is that when coarse-graining the edges, we want such a \emph{two-vertex interaction} to remain admissible at the coarse-grained level. This would not be the case if we were to coarse-grain to a single edge between the pair of vertices. The details and discussion of the dynamics, as well as the various effective interactions which we consider, will be presented in section \ref{sec:Effective Dynamics}.

\begin{figure}[H]
	\centering
	\begin{tikzpicture}[baseline=(v)]
		\node[circle,fill=black,draw=black,inner sep=1.5pt,label=below:$\iota_{v}$] (v) at (0,0) {};
		\node[circle,fill=black,draw=black,inner sep=0.7pt] (pl1) at (160:0.75) {};
		\node[circle,fill=black,draw=black,inner sep=0.7pt] (pl1) at (180:0.75) {};
		\node[circle,fill=black,draw=black,inner sep=0.7pt] (pl1) at (-160:0.75) {};
		\node[circle,fill=black,draw=black,inner sep=0.7pt] (pr1) at (20:0.75) {};
		\node[circle,fill=black,draw=black,inner sep=0.7pt] (pr1) at (0:0.75) {};
		\node[circle,fill=black,draw=black,inner sep=0.7pt] (pr1) at (-20:0.75) {};
		\draw[thick] (v) -- ++(135:1.5);
		\draw[thick] (v) -- ++(-135:1.5);
		\draw[thick] (v) -- ++(45:1.5);
		\draw[thick] (v) -- ++(-45:1.5);
	\end{tikzpicture}
	$\qquad\sim\qquad$
	\begin{tikzpicture}[baseline=(v)]
		\node[circle,fill=black,draw=black,inner sep=1.5pt,label=below:$\iota_{v}^{(1)}$] (v) at (0,0) {};
		\node[circle,fill=black,draw=black,inner sep=1.5pt,label=below:$\iota_{v}^{(2)}$] (vprime) at (1,0) {};
		\node[circle,fill=black,draw=black,inner sep=0.7pt] (pl1) at (160:0.75) {};
		\node[circle,fill=black,draw=black,inner sep=0.7pt] (pl1) at (180:0.75) {};
		\node[circle,fill=black,draw=black,inner sep=0.7pt] (pl1) at (-160:0.75) {};
		\node[circle,fill=black,draw=black,inner sep=0.7pt] (pr1) at ($(vprime)+(20:0.75)$) {};
		\node[circle,fill=black,draw=black,inner sep=0.7pt] (pr1) at ($(vprime)+(0:0.75)$) {};
		\node[circle,fill=black,draw=black,inner sep=0.7pt] (pr1) at ($(vprime)+(-20:0.75)$) {};
		\draw[thick] (v) -- ++(135:1.5);
		\draw[thick] (v) -- ++(-135:1.5);
		\draw[thick] (vprime) -- ++(45:1.5);
		\draw[thick] (vprime) -- ++(-45:1.5);
		\draw[thick,dashed] (v) -- (vprime) node[midway,above] {$k$} (v);
	\end{tikzpicture}
	$\qquad\longrightarrow\qquad$
	\begin{tikzpicture}[baseline=(v)]
		\node[circle,fill=black,draw=black,inner sep=1.5pt,label=below:$\iota_{v}$] (v) at (0,0) {};
		\node[circle,fill=black,draw=black,inner sep=1.5pt,label=below:$\iota_{v^\prime}$] (vprime) at (1,0) {};
		\node[circle,fill=black,draw=black,inner sep=0.7pt] (pl1) at (160:0.75) {};
		\node[circle,fill=black,draw=black,inner sep=0.7pt] (pl1) at (180:0.75) {};
		\node[circle,fill=black,draw=black,inner sep=0.7pt] (pl1) at (-160:0.75) {};
		\node[circle,fill=black,draw=black,inner sep=0.7pt] (pr1) at ($(vprime)+(20:0.75)$) {};
		\node[circle,fill=black,draw=black,inner sep=0.7pt] (pr1) at ($(vprime)+(0:0.75)$) {};
		\node[circle,fill=black,draw=black,inner sep=0.7pt] (pr1) at ($(vprime)+(-20:0.75)$) {};
		\draw[thick] (v) -- ++(135:1.5);
		\draw[thick] (v) -- ++(-135:1.5);
		\draw[thick] (vprime) -- ++(45:1.5);
		\draw[thick] (vprime) -- ++(-45:1.5);
		\draw[thick] (v) -- node[midway,above] {$\mathbb{1}^{(k)}$} (vprime);
	\end{tikzpicture}
	\vspace*{5mm}
	\caption{Illustrating the procedure to split a coarse vertex through the decomposition of the intertwiner. The internal spin $k$ (dashed line) is promoted to a trivial holonomy (solid line).}
	\label{fig:vertex_splitting}
\end{figure}
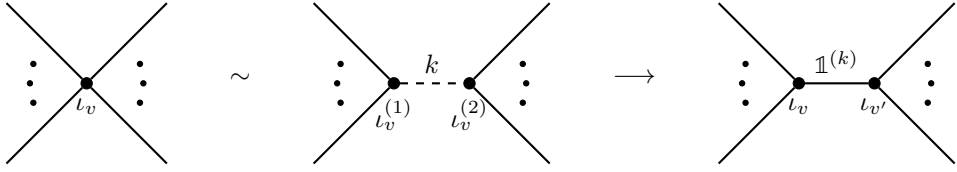

The core idea to coarse-grain edges is to use a decomposition of the intertwiners at the coarse vertices in order to create new edges and collapse the pre-existing edges to self-loops at these vertices. As illustrated in \figref{fig:vertex_splitting}, this \emph{splitting} of a vertex $v$ consists of choosing a spin recoupling basis for the intertwiner, where the spins of the edges to be collapsed to self-loops are grouped together, then recoupled to the remaining spins through a set of internal spins that partially characterizes the basis elements. We then promote these internal spins to trivial holonomies, thereby temporarily creating new vertices with intertwiners obtained as
\begin{align}
	&(\iota_{v})\indices{_{\mu \ldots}^{\nu \ldots}} = \sum_b c_b\ (\iota_{v}^{(b)})\indices{_{\mu \ldots}^{\nu \ldots}} = \sum_b c_b\ (\iota_{v}^{(b,1)})_{\mu_1\ldots} \,\, \delta^{\mu_1}_{\nu_1}[k_1] \,\, (\iota_{v}^{(b,2)})^{\,\nu_1\ldots}\, \dots \, \delta^{\mu_i}_{\nu_i}[k_i] \,\, (\iota_{v}^{(b,i)})^{\,\nu_i\ldots}\\
	&(\iota_{v}^{(b,1)})_{\mu_1\ldots} \,\, \delta^{\mu_1}_{\nu_1}[k_1] \,\, (\iota_{v}^{(b,2)})^{\,\nu_1\ldots}\, \dots
	\qquad\longrightarrow\qquad (\iota_v)_{\mu_1 \ldots} \,\, \qty[\mathbb{1}^{(k_1)}]\indices{^{\mu_1}_{\nu_1}} \,\, (\iota_{v^\prime})^{\,\nu_1 \ldots} \dots \ ,
\end{align}
where the intertwiner $\iota_{v}$ is decomposed in the chosen recoupling basis elements $(\iota_{v}^{(b)})$, with components $(\iota_{v}^{(b,i)})$, representing the coupling of the chosen groups of edges at the vertex, and contracted at the internal spins $k_i$. These are then promoted to the identity group elements.

We now develop the procedure to coarse-grain edges. Consider two vertices $v$ and $\tilde v$, with sets of edges $(\mathfrak e_v,\mathfrak e_{\tilde v})$ and valences $(w_v,w_{\tilde v})$ respectively, connected by $w$ edges $\{e_1,\dots,e_w\}$ with spins $j_i$ and holonomies $h_{e_i}^{(j_i)}$. The coarse-graining of edges to $p < w$ is realized as follows:
\begin{enumerate}
	\itemsep0em
	
	\item Partition the edges into $p$ disjoint sets $\mathfrak X_i$, allowing up to $p-1$ sets to be empty, such that $\cup_{i=1}^{p}\mathfrak X_i = \{e_1,\dots,e_w\}$. Assign $q$ of these sets to the vertex $v$ and the remaining sets to the vertex $\tilde v$. Denote them $\mathfrak X_i^v$ with cardinality $x_i^v$ and $\mathfrak X_i^{\tilde v}$ with cardinality $x_i^{\tilde v}$, respectively. 
	The assignment of sets must satisfy $w_v - \sum_{i=q+1}^p x_i^{\tilde v} \geq 2$ and $w_{\tilde v} - \sum_{i=1}^q x_i^v \geq 2$. This ensures that neither vertex is reduced to a trivial vertex, i.e.~a vertex with valence one or two.
	For instance, associate the first $q$ sets $\mathfrak X_i^v=\{e_{m_i},\dots,e_{m_{i+1}-1}\}$ to the vertex $v$ and the rest of the sets $\mathfrak X_i^{\tilde v}$ to the vertex $\tilde v$, such that $m_1 = 1$ and $m_{p+1} -1 = w$. The self-loops which will eventually be generated at the vertex $v$ correspond to edges in the sets $\mathfrak X_i^v$; and similarly for the vertex $\tilde v$.
	
	\item Apply the vertex splitting procedure described above to each vertex separately, such that the intertwiner decomposition is dictated by the sets of edges associated to each. Namely, split the vertex $v$ by coupling the spins of the edges in each set $\mathfrak X_i^{\tilde v}$ together; then these sets are successively recoupled until all the sets associated to $\tilde v$ are exhausted; and finally these sets are coupled to the remaining edges at $v$. Proceed similarly for the vertex $\tilde v$. Such a decomposition is depicted in the figure below, where the splitting of the two vertices is characterized by internal spins $k_i$:
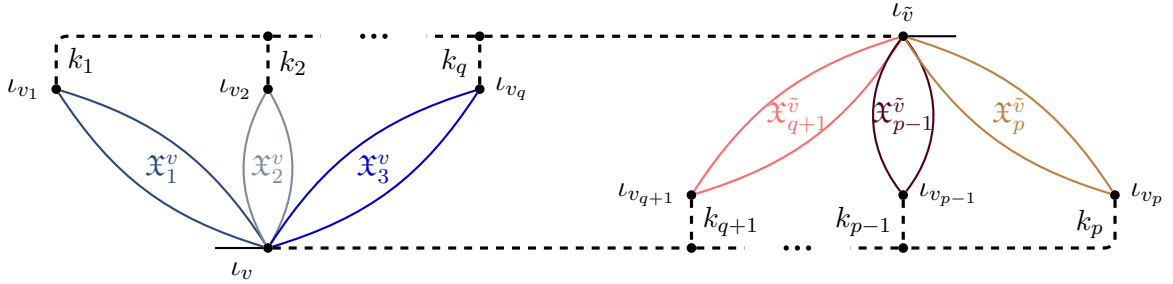
	\begin{center}
		\vspace{-0,5cm}
			\begin{tikzpicture}[thick,baseline=(v),scale=1.4]
				\node[circle,fill=black,draw=black,inner sep=1pt,label=below left:$\iota_{v}$] (v) at (0,0) {};
				\node[circle,fill=black,draw=black,inner sep=1pt,label=left:$\iota_{v_1}$] (vtilde1) at (-2,1.5) {};
				\node[circle,fill=black,draw=black,inner sep=1pt] (vtildeI) at (0,2) {};
				\node[circle,fill=black,draw=black,inner sep=1pt] (vtildeQ-1) at (2,2) {};
				\node[circle,fill=black,draw=black,inner sep=1pt,label=left:$\iota_{v_2}$] (vtilde2) at (0,1.5) {};
				\node[circle,fill=black,draw=black,inner sep=1pt,label=right:$\iota_{v_q}$] (vtildeq) at (2,1.5) {};
				\node[circle,fill=black,draw=black,inner sep=1pt,label=above:$\iota_{\tilde v}$] (vtilde) at (6,2) {};
				\node[circle,fill=black,draw=black,inner sep=1pt,label=right:$\iota_{v_p}$] (vp) at (8,0.5) {};
				\node[circle,fill=black,draw=black,inner sep=1pt] (vP-1) at (6,0) {};
				\node[circle,fill=black,draw=black,inner sep=1pt] (vQ) at (4,0) {};
				\node[circle,fill=black,draw=black,inner sep=1pt,label=right:$\iota_{v_{p-1}}$] (vp-1) at (6,0.5) {};
				\node[circle,fill=black,draw=black,inner sep=1pt,label=left:$\iota_{v_{q+1}}$] (vq+1) at (4,0.5) {};
				\draw (v) -- ++(-0.5,0);
				\draw (vtilde) -- ++(0.5,0);
				\draw[duskblue] (v) to[bend left=20] coordinate[pos=0.5] (1mid1) (vtilde1);
				\draw[duskblue] (v) to[bend right=20] coordinate[pos=0.5] (1mid2) (vtilde1);
				\node[duskblue] (X^v_1)  at ($(1mid1)!0.5!(1mid2)$) {$\mathfrak{X}^v_1$};
				\draw[lavendergrey] (v) to[bend left=30] coordinate[pos=0.5] (2mid1) (vtilde2);
				\draw[lavendergrey] (v) to[bend right=30] coordinate[pos=0.5] (2mid2) (vtilde2);
				\node[lavendergrey] (X^v_2)  at ($(2mid1)!0.5!(2mid2)$) {$\mathfrak{X}^v_2$};
				\draw[blue!80!black] (v) to[bend left=20] coordinate[pos=0.5] (3mid1) (vtildeq);
				\draw[blue!80!black] (v) to[bend right=20] coordinate[pos=0.5] (3mid2) (vtildeq);
				\node[blue!80!black] (X^v_3)  at ($(3mid1)!0.5!(3mid2)$) {$\mathfrak{X}^v_3$};
				\threedots{$(vtildeI)!0.5!(vtildeQ-1)$}
				\threedots{$(vQ)!0.5!(vP-1)$}
				\draw[grapefruitpink] (vq+1) to[bend left=20] coordinate[pos=0.5] (4mid1)(vtilde);
				\draw[grapefruitpink] (vq+1) to[bend right=20] coordinate[pos=0.5] (4mid2)	(vtilde);
				\node[grapefruitpink] (X^vtilde_q+1)  at ($(4mid1)!0.5!(4mid2)$) {$\mathfrak{X}^{\tilde{v}}_{q+1}$};
				\draw[crimsonviolet] (vp-1) to[bend left=38] coordinate[pos=0.5] (5mid1) (vtilde);
				\draw[crimsonviolet] (vp-1) to[bend right=38] coordinate[pos=0.5] (5mid2) (vtilde);
				\node[crimsonviolet] (X^vtilde_p-1)  at ($(5mid1)!0.5!(5mid2)$) {$\mathfrak{X}^{\tilde{v}}_{p-1}$};
				\draw[brown] (vp) to[bend left=20] coordinate[pos=0.5] (6mid1) (vtilde);
				\draw[brown] (vp) to[bend right=20] coordinate[pos=0.5] (6mid2) (vtilde);
				\node[brown] (X^vtilde_p)  at ($(6mid1)!0.5!(6mid2)$) {$\mathfrak{X}^{\tilde{v}}_p$};
				\draw[very thick,dashed,rounded corners] (vtilde1) -- node[pos=0.5,right]{$k_1$} ($(vtilde1)+(0,0.5)$) -- (vtildeI);
				\draw[very thick,dashed] (vtilde2) -- node[midway,right]{$k_2$} (vtildeI);
				\draw[very thick,dashed] (vtildeq) -- node[midway,left]{$k_q$} (vtildeQ-1);
				\draw[very thick,dashed] (vtildeQ-1) -- (vtilde);
				\draw[very thick,dashed] (v) -- (vQ);
				\draw[very thick,dashed] (vq+1) -- node[midway,right]{$k_{q+1}$} (vQ);
				\draw[very thick,dashed] (vp-1) -- node[midway,left]{$k_{p-1}$} (vP-1);
				\draw[very thick,dashed,rounded corners] (vp) -- node[pos=0.5,left]{$k_p$} ($(vp)-(0,0.5)$) -- (vP-1);
				\draw[very thick, dashed] (vtildeI) -- ++(0.5,0);
				\draw[very thick, dashed] (vtildeQ-1) -- ++(-0.5,0);	
				\draw[very thick, dashed] (vQ) -- ++(0.5,0);
				\draw[very thick, dashed] (vP-1) -- ++(-0.5,0);
			\end{tikzpicture}
	\end{center}
	The next steps require distinguishing the cases where one of the internal spins $k_i$ vanishes, from the cases where none of them vanishes. This is because promoting an internal spin zero to a group element produces an edge with spin zero, and such an edge is always trivial and can be dropped from the state by cylindrical consistency. However, removing an edge from the graph would automatically reduce the number of independent loops, which in turn would violate our requirement of preserving the number of independent loops under coarse-graining. To resolve this, one must observe that every intertwiner state is mapped to many coarse states, each induced by a different choice of the sets $\mathfrak X_i$. When one of the internal spins given by the vertex splitting of an intertwiner vanishes, one simply has to use a different partition of edges to coarse-grain this particular state. The only exception is when only one set $\mathfrak X_i$ is non-empty, as in this case the splitting at vanishing internal spin does not change the number of independent loops.
	\newpage
	\item Assuming that none of the internal spins $k_i$ resulting from the splitting vanishes, promote these internal spins to trivial group elements:
	
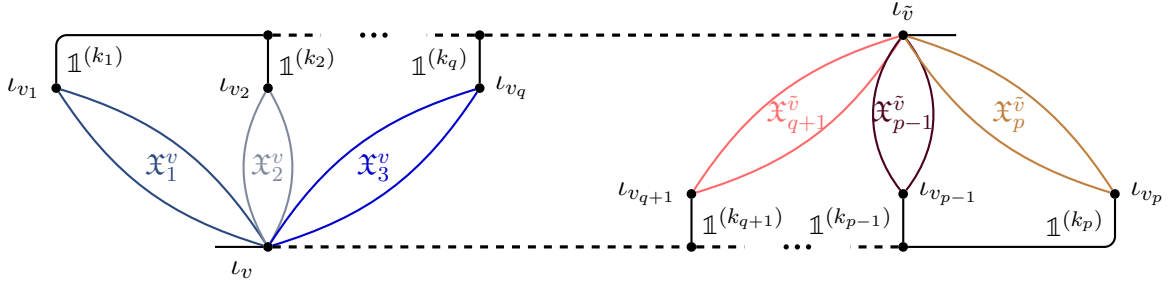
\begin{center}
		\vspace{-0,5cm}
			\begin{tikzpicture}[thick,baseline=(v),scale=1.4]
				\node[circle,fill=black,draw=black,inner sep=1pt,label=below left:$\iota_{v}$] (v) at (0,0) {};
				\node[circle,fill=black,draw=black,inner sep=1pt,label=left:$\iota_{v_1}$] (vtilde1) at (-2,1.5) {};
				\node[circle,fill=black,draw=black,inner sep=1pt] (vtildeI) at (0,2) {};
				\node[circle,fill=black,draw=black,inner sep=1pt] (vtildeQ-1) at (2,2) {};
				\node[circle,fill=black,draw=black,inner sep=1pt,label=left:$\iota_{v_2}$] (vtilde2) at (0,1.5) {};
				\node[circle,fill=black,draw=black,inner sep=1pt,label=right:$\iota_{v_q}$] (vtildeq) at (2,1.5) {};
				\node[circle,fill=black,draw=black,inner sep=1pt,label=above:$\iota_{\tilde v}$] (vtilde) at (6,2) {};
				\node[circle,fill=black,draw=black,inner sep=1pt,label=right:$\iota_{v_p}$] (vp) at (8,0.5) {};
				\node[circle,fill=black,draw=black,inner sep=1pt] (vP-1) at (6,0) {};
				\node[circle,fill=black,draw=black,inner sep=1pt] (vQ) at (4,0) {};
				\node[circle,fill=black,draw=black,inner sep=1pt,label=right:$\iota_{v_{p-1}}$] (vp-1) at (6,0.5) {};
				\node[circle,fill=black,draw=black,inner sep=1pt,label=left:$\iota_{v_{q+1}}$] (vq+1) at (4,0.5) {};
				\draw (v) -- ++(-0.5,0);
				\draw (vtilde) -- ++(0.5,0);
				\draw[duskblue] (v) to[bend left=20] coordinate[pos=0.5] (1mid1) (vtilde1);
				\draw[duskblue] (v) to[bend right=20] coordinate[pos=0.5] (1mid2) (vtilde1);
				\node[duskblue] (X^v_1)  at ($(1mid1)!0.5!(1mid2)$) {$\mathfrak{X}^v_1$};
				\draw[lavendergrey] (v) to[bend left=30] coordinate[pos=0.5] (2mid1) (vtilde2);
				\draw[lavendergrey] (v) to[bend right=30] coordinate[pos=0.5] (2mid2) (vtilde2);
				\node[lavendergrey] (X^v_2)  at ($(2mid1)!0.5!(2mid2)$) {$\mathfrak{X}^v_2$};
				\draw[blue!80!black] (v) to[bend left=20] coordinate[pos=0.5] (3mid1) (vtildeq);
				\draw[blue!80!black] (v) to[bend right=20] coordinate[pos=0.5] (3mid2) (vtildeq);
				\node[blue!80!black] (X^v_3)  at ($(3mid1)!0.5!(3mid2)$) {$\mathfrak{X}^v_3$};
				\threedots{$(vtildeI)!0.5!(vtildeQ-1)$}
				\threedots{$(vQ)!0.5!(vP-1)$}
				\draw[grapefruitpink] (vq+1) to[bend left=20] coordinate[pos=0.5] (4mid1)(vtilde);
				\draw[grapefruitpink] (vq+1) to[bend right=20] coordinate[pos=0.5] (4mid2)	(vtilde);
				\node[grapefruitpink] (X^vtilde_q+1)  at ($(4mid1)!0.5!(4mid2)$) {$\mathfrak{X}^{\tilde{v}}_{q+1}$};
				\draw[crimsonviolet] (vp-1) to[bend left=38] coordinate[pos=0.5] (5mid1) (vtilde);
				\draw[crimsonviolet] (vp-1) to[bend right=38] coordinate[pos=0.5] (5mid2) (vtilde);
				\node[crimsonviolet] (X^vtilde_p-1)  at ($(5mid1)!0.5!(5mid2)$) {$\mathfrak{X}^{\tilde{v}}_{p-1}$};
				\draw[brown] (vp) to[bend left=20] coordinate[pos=0.5] (6mid1) (vtilde);
				\draw[brown] (vp) to[bend right=20] coordinate[pos=0.5] (6mid2) (vtilde);
				\node[brown] (X^vtilde_p)  at ($(6mid1)!0.5!(6mid2)$) {$\mathfrak{X}^{\tilde{v}}_p$};
				\draw[thick,rounded corners] (vtilde1) -- node[pos=0.5,right]{$\mathbb{1}^{(k_1)}$} ($(vtilde1)+(0,0.5)$) -- (vtildeI);
				\draw[thick] (vtilde2) -- node[midway,right]{$\mathbb{1}^{(k_2)}$} (vtildeI);
				\draw[thick] (vtildeq) -- node[midway,left]{$\mathbb{1}^{(k_q)}$} (vtildeQ-1);
				\draw[very thick,dashed] (vtildeQ-1) -- (vtilde);
				\draw[very thick,dashed] (v) -- (vQ);
				\draw[thick] (vq+1) -- node[midway,right]{$\mathbb{1}^{(k_{q+1})}$} (vQ);
				\draw[thick] (vp-1) -- node[midway,left]{$\mathbb{1}^{(k_{p-1})}$} (vP-1);
				\draw[thick,rounded corners] (vp) -- node[pos=0.5,left]{$\mathbb{1}^{(k_p)}$} ($(vp)-(0,0.5)$) -- (vP-1);
				\draw[very thick, dashed] (vtildeI) -- ++(0.5,0);
				\draw[very thick, dashed] (vtildeQ-1) -- ++(-0.5,0);	
				\draw[very thick, dashed] (vQ) -- ++(0.5,0);
				\draw[very thick, dashed] (vP-1) -- ++(-0.5,0);
			\end{tikzpicture}
	\end{center}
	effectively creating at most $p$ new vertices $v_i$ with $p$ new edges as shown above. The exact number is actually equal to the number of non-empty sets $\mathfrak X_i$ in the chosen partition. Note that if a set $\mathfrak X_i$ contains only one edge, then this step and the following one are unnecessary because such an edge is untouched and is mapped to itself as a coarse edge.
	
	\item Apply the gauge fixing procedure to coarse-grain vertices and collapse the vertices $v_i^v \equiv v_i$ for $1\leq i \leq q$ into $v$, and the vertices $v_i^{\tilde v} \equiv v_i$ for $q+1\leq i \leq p$ into $\tilde v$. First select an edge in each set $\mathfrak X_i^v$, say $e_{m_i}$. Second, gauge fix the holonomies $h_{e_{m_i}}$ to the identity group element by making gauge transformations at the associated vertices $v_i$. Finally, collapse these vertices $v_i^v$ into $v$ such that the edges in each $\mathfrak X_i^v\backslash \{e_{m_i}\}$ become self-loops at $v$. Proceed similarly to collapse the vertices $v_i^{\tilde v}$ into $\tilde v$. One then arrives at the following configuration:
	
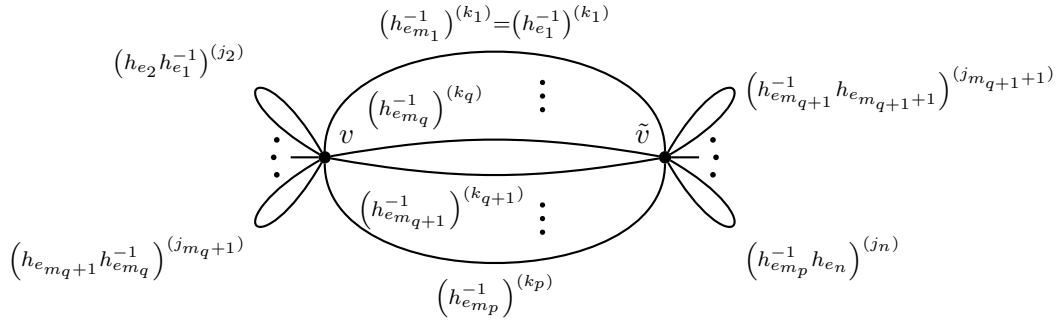
\begin{center}
		\begin{tikzpicture}[baseline=(v),scale=0.9]
			\node[circle,fill=black,draw=black,inner sep=1.5pt,label=above right:$v$] (v) at (0,0) {};
			\node[circle,fill=black,draw=black,inner sep=1.5pt,label=above left:$\tilde v$] (vtilde) at (5,0) {};
			\draw[thick] (v) -- ++(-0.5,0);
			\draw[thick] (vtilde) -- ++(0.5,0);
			\draw[thick] (v) to[out=90,in=90] node[midway,above]{${\scriptstyle \qty(h_{e_{m_1}}^{-1})^{(k_1)}=\qty(h_{e_1}^{-1})^{(k_1)}}$} (vtilde);
			\draw[thick] (v) to[out=-90,in=-90] node[midway,below]{${\scriptstyle \qty(h_{e_{m_p}}^{-1})^{(k_{p})}}$} (vtilde);
			\draw[thick] (v) to[out=10,in=170] node[midway,above left]{${\scriptstyle \qty(h_{e_{m_q}}^{-1})^{(k_q)}}$} (vtilde);
			\draw[thick] (v) to[out=-10,in=-170] node[midway, below, xshift=-20]{${\scriptstyle \qty(h_{e_{m_{q+1}}}^{-1})^{(k_{q+1})}}$} (vtilde);
			\node[circle,fill,inner sep=0.7pt] (icg) at ($($(v)!0.5!(vtilde)$)+(0.7,1.1)$) {};
			\node[circle,fill,inner sep=0.7pt] (iicg) at ($($(v)!0.5!(vtilde)$)+(0.7,0.9)$) {};
			\node[circle,fill,inner sep=0.7pt] (iiicg) at ($($(v)!0.5!(vtilde)$)+(0.7,0.7)$) {};
			\node[circle,fill,inner sep=0.7pt] (icg) at ($($(v)!0.5!(vtilde)$)+(0.7,-1.1)$) {};
			\node[circle,fill,inner sep=0.7pt] (iicg) at ($($(v)!0.5!(vtilde)$)+(0.7,-0.9)$) {};
			\node[circle,fill,inner sep=0.7pt] (iiicg) at ($($(v)!0.5!(vtilde)$)+(0.7,-0.7)$) {};
			\draw[thick,black] (v) to[out=120,in=150,looseness=100] node[midway,above left]{${\scriptstyle \qty(h_{e_2}h_{e_1}^{-1})^{(j_{2})}}$} (v);
			\node[circle,fill=black,inner sep=0.75pt] at ($(v)+(20:-0.75)$) {};
			\node[circle,fill=black,inner sep=0.75pt] at ($(v)+(0:-0.75)$) {};
			\node[circle,fill=black,inner sep=0.75pt] at ($(v)+(-20:-0.75)$) {};
			\draw[thick,black] (v) to[out=-120,in=-150,looseness=100] node[midway,below left]{${\scriptstyle \qty(h_{e_{m_q+1}}h_{e_{m_q}}^{-1})^{(j_{m_q+1})}}$} (v);
			\draw[thick,black] (vtilde) to[out=30,in=60,looseness=100] node[midway,above,right]{${\scriptstyle \qty(h_{e_{m_{q+1}}}^{-1}h_{e_{m_{q+1}+1}})^{(j_{m_{q+1}+1})}}$} (vtilde);
			\node[circle,fill=black,inner sep=0.75pt] at ($(vtilde)+(20:0.75)$) {};
			\node[circle,fill=black,inner sep=0.75pt] at ($(vtilde)+(0:0.75)$) {};
			\node[circle,fill=black,inner sep=0.75pt] at ($(vtilde)+(-20:0.75)$) {};
			\draw[thick,black] (vtilde) to[out=-30,in=-60,looseness=100] node[midway,below right]{${\scriptstyle \qty(h_{e_{m_{p}}}^{-1}h_{e_n})^{(j_{n})}}$} (vtilde);
		\end{tikzpicture}
	\end{center}
\end{enumerate}
This procedure ensures that the curvature which was encoded in the loops formed by the multiple edges is systematically encoded in self-loops distributed between the two vertices, plus the loops formed by the coarse edges. 
For clarity, we illustrate this whole procedure in an example in Table \ref{tab:edge cg algorithm example}.
\begin{table}[b]
	\caption{The edge coarse-graining algorithm applied to a pair of vertices connected by four edges.}
	\label{tab:edge cg algorithm example}
	\setlength{\tabcolsep}{1em}
	\centering
	\begin{tabular}{c|c|c|c}
		\toprule
		1. choose a partition	&	2. split the vertices	&   3. promote $k_i$ to $\mathbb{1}^{(k_i)}$	&	 4. gauge fix + collapse \\
		\midrule
		\begin{tikzpicture}[thick,baseline=(v)]
			\node[p_node] (v) at (0,0) {};
			\node[p_node] (tildev) at (2,0) {};
			\draw (v) -- ++(-0.25,0.25);
			\draw (v) -- ++(-0.25,-0.25);
			\draw (tildev) -- ++(0.25,0.25);
			\draw (tildev) -- ++(0.25,-0.25);
			\draw[duskblue] (v) to[out=60,in=120] node[midway,above]{$\mathfrak{X}^{v}_1 =\{e_1,e_2\}$} (tildev);
			\draw[duskblue] (v) to[out=30,in=150] (tildev);
			\draw[grapefruitpink] (v) to[out=-30,in=-150] (tildev);
			\draw[grapefruitpink] (v) to[out=-60,in=-120] node[midway,below]{$\mathfrak{X}^{\tilde{v}}_2=\{e_3,e_4\}$} (tildev);
		\end{tikzpicture} 
		&
		\begin{tikzpicture}[thick,baseline=($(v)!0.5!(tildev)$)]
			\node[p_node] (v) at (0,0) {};
			\node[p_node] (v1) at (1,1) {};
			\node[p_node] (v2) at (1,0) {};
			\node[p_node] (tildev) at (2,1) {};
			\draw (v) -- ++(-0.25,0.25);
			\draw (v) -- ++(-0.25,-0.25);
			\draw (tildev) -- ++(0.25,0.25);
			\draw (tildev) -- ++(0.25,-0.25);
			\draw[dashed] (v) -- node[midway,below]{$k_2$} (v2);
			\draw[dashed] (v1) -- node[midway,above]{$k_1$} (tildev);
			\draw[duskblue] (v) to[out=60,in=-150] (v1);
			\draw[duskblue] (v) to[out=30,in=-120] (v1);
			\draw[grapefruitpink] (v2) to[out=60,in=-150] (tildev);
			\draw[grapefruitpink] (v2) to[out=30,in=-120] (tildev);
		\end{tikzpicture}
		&
		\begin{tikzpicture}[thick,baseline=($(v)!0.5!(tildev)$)]
			\node[p_node] (v) at (0,0) {};
			\node[p_node] (v1) at (1,1) {};
			\node[p_node] (v2) at (1,0) {};
			\node[p_node] (tildev) at (2,1) {};
			\draw (v) -- ++(-0.25,0.25);
			\draw (v) -- ++(-0.25,-0.25);
			\draw (tildev) -- ++(0.25,0.25);
			\draw (tildev) -- ++(0.25,-0.25);
			\draw (v) -- node[midway,below]{$\mathbb{1}^{(k_2)}$} (v2);
			\draw (v1) -- node[midway,above]{$\mathbb{1}^{(k_1)}$} (tildev);
			\draw[duskblue] (v) to[out=60,in=-150] (v1);
			\draw[duskblue] (v) to[out=30,in=-120] (v1);
			\draw[grapefruitpink] (v2) to[out=60,in=-150] (tildev);
			\draw[grapefruitpink] (v2) to[out=30,in=-120] (tildev);
		\end{tikzpicture}
		&
		\begin{tikzpicture}[thick,baseline=(v)]
			\node[p_node] (v) at (0,0) {};
			\node[p_node] (tildev) at (2,0) {};
			\draw (v) -- ++(-0.25,0.25);
			\draw (v) -- ++(-0.25,-0.25);
			\draw (tildev) -- ++(0.25,0.25);
			\draw (tildev) -- ++(0.25,-0.25);
			\draw[duskblue] (v) to[out=110,in=70,looseness=40] (v);
			\draw[grapefruitpink] (tildev) to[out=110,in=70,looseness=40] (tildev);
			\draw (v) to[out=45,in=135] node[midway,above]{$\qty(h^{-1}_{e_1})^{(k_1)}$} (tildev);
			\draw (v) to[out=-45,in=-135] node[midway,below]{$\qty(h^{-1}_{e_3})^{(k_2)}$} (tildev);
		\end{tikzpicture} \\
		\bottomrule
	\end{tabular}
	\setlength{\tabcolsep}{-2em}
\end{table}

Some observations regarding the above procedure are in order. First, notice that indeed if one $k_i$ were equal to zero, the outcome of step 4.~above, in the case of more than one non-empty set $\mathfrak X_i$, would be that starting with $w-1$ independent loops formed by the edges $\{e_1,\dots,e_w\}$, we arrive at $w-2$ loops after coarse-graining. Hence, the number of independent loops would not be preserved. This is why the splitting cannot occur at internal spin zero unless the partition consists of a single non-empty set.
Second, in step 1.~we imposed certain conditions on the partition in order to guarantee that the procedure does not result in a reduction of the number of vertices. For a fixed $p$, there may exist states which contain a pair of vertices that do not admit a partition satisfying these conditions. This would mean that for such a particular pair, the number of edges cannot be reduced to $p$ coarse edges. An example is the \emph{Theta-graph} state (two three-valent vertices) when $p=2$. At present, we see two solutions for such cases: either we relax the requirement of not reducing the number of vertices while coarse-graining the edges; or we impose that such states are mapped to the zero state via this procedure. For all that follows in this article, it is not relevant which option we choose, but we expect that the differences primarily manifest at the computational level, for instance when solving the renormalization flow equations.
Finally, a coarse state resulting from this procedure has a number of coarse edges which is equal to the number of non-empty sets $\mathfrak X_i$; except for the partition given by one non-empty set, in which case it is possible to have no coarse edge at all if the splitting happens at spin equal to zero. Generally, for a partition with $z$ empty sets, the induced mapping can be viewed as generating $z$ coarse-edges with spins equal to zero, while preserving the number of independent loops. The inclusion of such partitions of edges in our procedure guarantees that all intertwiners have non-trivial images (not zero). Additionally,  as noted earlier, a set $\mathfrak X_i$ containing only one edge is trivially coarse-grained to itself.

Obviously, this procedure for the edges requires making certain choices: the subsets $\mathfrak X_i^v$ and $\mathfrak X_i^{\tilde v}$, and the choice of edges whose holonomies are trivialized (which is basically the choice of trees). But as already mentioned for the coarse-graining of vertices, such dependencies can be removed via an averaging over the finite number of choices. We discuss this aspect in detail in section \ref{sec:CG maps and spaces}.

Now that we have described how to coarse-grain vertices and edges, the last element in our procedure is the coarse-graining of geometric tags for the states in $\mathcal H_{\rm T}$. This is the subject of the next section.


\subsection{Coarse-graining geometric tags}\label{sec:coarse-graining tags}

We examine two ideas for the elaboration of a coarse-graining procedure for the geometric tags.
The first method is straightforward and it consists of keeping the geometric tag associated to each vertex of the initial state intact. Namely, when coarse-graining vertices, the resulting coarse vertex will carry a geometric tag which is simply the union of all the geometric tags associated to the vertices of the maximal tree, without the elements involving the edges of the tree, then completed with zeros associated to any pair or triple of edges which involve different initial vertices. 

To illustrate this proposal, consider the example of two four-valent vertices $v$ and $\tilde v$, with sets of edges $\mathfrak{e}_v = \{e_1,e_2,e_3,e_4\}$ and $\mathfrak{e}_{\tilde v} = \{e_3,e_4,e_5,e_6\}$ respectively. Say we coarse-grain the two vertices to one vertex $v_o$ by trivializing the holonomy along $e_4$. Since the edge $e_3$ becomes a loop at $v_o$, we can effectively split the edge into two segments in order to keep track of the different edges at $v_o$. Let us denote by $e_3^v$ the segment associated to $v$ and by $e_3^{\tilde v}$ the segment associated to $\tilde v$. These segments, $e_3^v$ and $e_3^{\tilde v}$, replace $e_3$ in $\mathfrak{e}_v$ and $\mathfrak{e}_{\tilde v}$ respectively, and we have the set of edges $\mathfrak{e}_v^o = \{e_1,e_2,e_3^v, e_3^{\tilde v},e_5, e_6\}$ at $v_o$. Then the above method generates the following mappings for the coarse vertex $v_o$
\begin{alignat}{2}
	\vartheta_{v_o}(e_i, e_j) = 
	\begin{cases}
		\vartheta_{v}(e_i, e_j)\ ,\ \text{if $e_i, e_j \in \mathfrak{e}_v$} \\ \vartheta_{\tilde v}(e_i, e_j)\ ,\ \text{if $e_i, e_j \in \mathfrak{e}_{\tilde v}$} \\ 0\ ,\ \text{otherwise}
	\end{cases}
	, \quad
	\varepsilon_{v_o}(e_i, e_j, e_k) = \begin{cases}
		\varepsilon_{v}(e_i, e_j, e_k)\ ,\ \text{if $e_i, e_j, e_k \in \mathfrak{e}_v$} \\ \varepsilon_{\tilde v}(e_i, e_j, e_k)\ ,\ \text{if $e_i, e_j, e_k \in \mathfrak{e}_{\tilde v}$} \\ 0\ ,\ \text{otherwise}
	\end{cases}
	\label{CG_Geo_Tags_1}
\end{alignat}
where the maps $\vartheta$ and $\varepsilon$ are the ones defined in \eqref{Geo_Tags}. 
The mappings in \eqref{CG_Geo_Tags_1} induce a very minimal coarse-graining of the geometric tags. Namely, only the data involving the edges of the maximal tree are lost (edge $e_4$ in this example). Also, the structure of the geometric tag at the coarse vertex permits tracing back the number of vertices which were collapsed, with all their algebraic data except for the edges of the maximal tree. This is too much data to be retained, as it significantly undermines the purpose of coarse-graining vertices. We hence propose a second approach which does not compromise the coarse-graining of the combinatorial and algebraic data presented previously.

Indeed, instead of prohibiting any mixing of the edges from different initial vertices in the geometric data at the coarse vertex, the idea is to choose non-trivial geometric data associated to pairs and triples of edges which did not belong to the same initial vertex. In the example above, the maps in \eqref{CG_Geo_Tags_1} would not systematically give a zero value for the cases where $e_i$ and $e_j$ do not belong to the same set of edges ($\mathfrak{e}_v$ or $\mathfrak{e}_{\tilde v}$). Rather, we would have
\begin{alignat}{2}
	\vartheta_{v_o}(e_i, e_j) = 
	\begin{cases}
		\vartheta_{v}(e_i, e_j)\ ,\ \text{if $e_i, e_j \in \mathfrak{e}_v$} \\ \vartheta_{\tilde v}(e_i, e_j)\ ,\ \text{if $e_i, e_j \in \mathfrak{e}_{\tilde v}$} \\ \theta_{ij} \in \{0,1\}\ ,\ \text{otherwise}
	\end{cases}
	, \quad
	\varepsilon_{v_o}(e_i, e_j, e_k) = \begin{cases}
		\varepsilon_{v}(e_i, e_j, e_k)\ ,\ \text{if $e_i, e_j, e_k \in \mathfrak{e}_v$} \\ \varepsilon_{\tilde v}(e_i, e_j, e_k)\ ,\ \text{if $e_i, e_j, e_k \in \mathfrak{e}_{\tilde v}$} \\ \eta_{ijk} \in \{-1,0,1\} \ ,\ \text{otherwise}
	\end{cases}.
	\label{CG_Geo_Tags_2}
\end{alignat}
The values $\theta_{ij}$ and $\eta_{ijk}$ which complete the geometric tag at the coarse vertex $v_o$ must satisfy three requirements. The first is that they must be chosen so that they are consistent with the data already provided by the maps $\vartheta_{v}$, $\vartheta_{\tilde v}$, $\varepsilon_{v}$ and $\varepsilon_{\tilde v}$. In other words, the orientations of all triples of edges are consistent, and in particular for any triple of edges $(e_i,e_j,e_k)$ at $v_o$, if $\vartheta_{v_o}(e_i, e_j) = 0 = \vartheta_{v_o}(e_i, e_k)$, then $\vartheta_{v_o}(e_j, e_k) = 0$. 
The second requirement accounts for the coarse-graining of the edges, by imposing that the coarse edge acquires the geometric data of the trivialized edge between the pair of vertices\footnote{An alternative to this requirement is to average over all admissible choices of geometric tags involving the new coarse edge.}, while the edges which become loops are processed according to \eqref{CG_Geo_Tags_2}. 
The third requirement is much more subtle and relates to what we call degenerate vertices. A degenerate vertex is either a two- or three-valent vertex, or a \emph{planar vertex} where for any triple of edges $(e_i,e_j,e_k)$, $\varepsilon(e_i, e_j, e_k) = 0$. Such vertices are trivially annihilated by the volume operator introduced in \cite{Ashtekar:1997fb} and given in \eqref{V_AL}.
If we were to coarse-grain a set of degenerate vertices without taking into account this aspect in coarse-graining geometric tags, we would be able to produce coarse states with non-vanishing volume from vertices with strictly zero volume. From a geometric perspective, this would mean that on a kinematical level the coarse-graining process could produce a non-degenerate effective metric from fundamentally degenerate configurations. This seems difficult to justify from a physical point of view. Therefore, we formulate the third requirement in coarse-graining geometric tags as the necessity to map a set of degenerate vertices to a degenerate coarse vertex, by systematically mapping them to a planar vertex, consequently excluding generating volume excitations from fully degenerate vertices. Altogether, these three requirements define the method that we adopt to coarse-grain the geometric tags.
Note that the third requirement can be implemented only if we use geometric tags, which means that if we coarse-grain states in the Hilbert space $\mathcal H_{\rm A}$, we would not be able to impose the condition that zero-volume configurations should not produce volume excitations under coarse-graining. This provides an additional argument for introducing geometric tags and preferring the Hilbert space $\mathcal H_{\rm T}$ over the space $\mathcal H_{\rm A}$. In Table \ref{tab:cg geom tags ex}, we provide a simple example of the application of the coarse-graining method for geometric tags.
\begin{table}[!t]
	\caption{Example of coarse-graining geometric tags: two vertices $v$ and $v'$ (on the left) are collapsed to one coarse vertex with two self-loops (on the right) by trivializing the edge $e_{3}$. We denote each edge connecting the two vertices with its unprimed or primed index to distinguish its two segments containing $v$ or $v'$ respectively. We also drop the vertex label in our notation of $\vartheta$ and $\varepsilon$.}
	\setlength{\tabcolsep}{3em}
	\centering
	\begin{tabular}{c|c}
		\toprule
		\begin{tikzpicture}[thick,scale=2,baseline=(v)]
			\node[p_node] (v) at (0,0) {};
			\node[p_node] (vtilde) at (1,0) {};
			\draw (v) to[out=90,in=90] node[pos=0.25,above]{$1^\prime$} node[pos=0.75,above]{$1$} (vtilde);
			\draw (v) to node[pos=0.25,below]{$2^\prime$} node[pos=0.75,below]{$2$} (vtilde);
			\draw (v) to[out=-90,in=-90] node[pos=0.25,below]{$3^\prime$} node[pos=0.75,below]{$3$} (vtilde);
			\draw (vtilde) -- node[midway,above]{$4$} ++(0.5,0);
		\end{tikzpicture}
		&	
		\begin{tikzpicture}[thick,scale=2,baseline=(v0)]
			\node[p_node] (v0) at (0,0) {};
			\draw (v0) to[out=135,in=45,looseness=40] node[pos=0.25,left]{$1^\prime$} node[pos=0.75,right]{$1$} (v0);
			\draw (v0) to[out=-135,in=-45,looseness=40] node[pos=0.25,left]{$2^\prime$} node[pos=0.75,right]{$2$} (v0);
			\draw (v0) -- node[pos=0.75,above]{$4$} ++(0.5,0);
		\end{tikzpicture}
		\\
		\midrule
		\multicolumn{2}{c}{$\vartheta(i,j)$} \\
		\midrule
		\makecell[lt]{
			$\vartheta(1,2)=\vartheta(1^{\prime},2^{\prime})=\vartheta(2,4)=1,$ \\[0.5em] 
			$\vartheta(1,4)=0,$ \\[0.5em]
			$\vartheta(1,3)=\vartheta(2,3)=\vartheta(3,4)=1,$  \\[0.5em]
			$\vartheta(1^{\prime},3^{\prime})=\vartheta(2^{\prime},3^{\prime})=1$
		}	&	
		\makecell[lt]{
			$\vartheta(1,2)=\vartheta(1^{\prime},2^{\prime})=\vartheta(2,4)=1,$ \\[0.5em] 
			$\vartheta(1,4)=0,$ \\[0.5em]
			$\vartheta(1,1^\prime)=\vartheta(1^\prime,4)=\vartheta(1^\prime,2)=1,$ \\[0.5em]
			$\vartheta(1,2^\prime)=\vartheta(2,2^\prime)=\vartheta(2^\prime,4)=1$
		}	\\
		\midrule
		\multicolumn{2}{c}{$\varepsilon(i,j,k)$} \\
		\midrule
		\makecell[lt]{
			$\varepsilon(1,2,4)=0,$\\[0.5em]
			$\varepsilon(1,3,4)=0,$ \\[0.5em]
			$\varepsilon(1,2,3)=\varepsilon(2,3,4)=\varepsilon(1^\prime,2^\prime,3^\prime)=1,$ \\[0.5em]
			$\varepsilon(1^\prime,2^\prime,3^\prime)=-1$ 
		}	&	
		\makecell[lt]{
			$\varepsilon(1,2,4)=0,$ \\[0.5em]
			$\varepsilon(1^\prime,1,2)=\varepsilon(1^\prime,1,2^\prime)=\varepsilon(2^\prime,2,1^\prime)=1$\\[0.5em]
			$\varepsilon(1^\prime,2^\prime,4)=\varepsilon(1^\prime,2,4)=1$, \\[0.5em]
			$\varepsilon(2^\prime,2,1)=\varepsilon(2^\prime,2,4)=-1$, \\[0.5em]
			$\varepsilon(1^\prime,1,4)=\varepsilon(1,2,4)=0$
		}	\\
		\bottomrule
	\end{tabular}
	\label{tab:cg geom tags ex}
	\setlength{\tabcolsep}{-1em}
\end{table}

Naturally, for each initial configuration there is a limited number of consistent choices $(\theta_{ij},\eta_{ijk})$, and each choice would provide a different geometric tag at $v_o$, and consequently a different state. This freedom creates an ambiguity in the coarse-graining of geometric tags. However, since the number of choices is always finite, this ambiguity can be removed by averaging over all consistent configurations of geometric tags. As we explain further in the following section, averaging over the different choices inherent to the coarse-graining procedures allows removing these ambiguities.

Now that we have introduced the coarse-graining procedures for the abstract graph structure, i.e.~vertices and edges, as well as the geometric tags, we can finally define the coarse-graining maps and provide the kinematical structure of the effective theories used to define our coarse-grained models. We describe their construction in detail in the next section.


\subsection{Coarse-graining maps and Hilbert spaces with \mbox{$(n,p)$-complete} graphs}\label{sec:CG maps and spaces}

We first summarize the coarse-graining procedure we presented so far. Starting from the kinematical Hilbert space $\mathcal{H}_{\rm G}$, or $\mathcal{H}_{\rm diff}$, we move to the Hilbert space of states with abstract graphs, either with or without geometric tags, that is, $\mathcal{H}_{\rm T}$ or $\mathcal{H}_{\rm A}$ respectively. We then use the method of coarse-graining via gauge-fixing introduced in \cite{Charles:2016xwc} to coarse-grain the vertices of spin network states to an arbitrary number $n$ of coarse vertices. After that, we coarse-grain the edges to an arbitrary number ${p}$ of coarse edges. Finally, we coarse-grain the geometric tags in case we start with states in $\mathcal{H}_{\rm T}$. This procedure is distinguished by the fact that it preserves the number of independent loops in gauge invariant spin network states.

As we emphasized in the introduction, our objective is to construct coarse-grained models with concrete kinematical structure and effective dynamics, in addition to coarse-graining maps adequate for the implementation of the Hamiltonian renormalization framework for these models. To do so, we begin by defining the basic coarse-graining maps which map the states in $\mathcal{H}_{\rm T}$ or $\mathcal{H}_{\rm A}$ to the coarse states. We then identify the coarse Hilbert spaces, which constitute the spaces of the coarse states, and show that they form subspaces of the initial Hilbert space.

Similarly to the decomposition \eqref{decomp} of the Hilbert space $\mathcal H_{\rm kin}$, the Hilbert space $\mathcal{H}_{\rm T}$ decomposes as
\begin{align}\label{decomp_HT}
	{\mathcal H}_{\rm T} = \bigoplus_{\Gamma} {\tilde{\mathcal H}}_{\Gamma}\ ,
\end{align}
where the direct sum is over all abstract graphs with geometric tags.
We then introduce a family of linear maps $\mathcal K_{\Gamma}^{(n,p)}$ on each space ${\tilde{\mathcal H}}_{\Gamma}$ labeled by two parameters $n,\, p \in \mathbbm N^+ \cup \{\infty\}$, and which map the states in ${\tilde{\mathcal H}}_{\Gamma}$ to a subspace of ${\mathcal H}_{\rm T}$ by application of the procedures presented above to coarse-grain vertices, edges and geometric tags.

Each map $\mathcal K_{\Gamma}^{(n,p)}$ takes several parameters as arguments:
\begin{itemize}
	\itemsep0em

	\item a set of $n$ disjoint subgraphs $\{\gamma_i\}$ such that $\bigcup\limits_{i=1}^n \gamma_i = \Gamma$ (sec.~\ref{sec:coarse-graining vertices});
	
	\item a set of $n$ maximal trees $\{T_i\}$ with root vertices $v_i$, each in a subgraph $\gamma_i$ (sec.~\ref{sec:coarse-graining vertices});
	
	\item for each pair of distinct subgraphs $(\gamma_i,\gamma_j)$: $p$ sets $\mathfrak X_c$ of edges such that $\cup_c \mathfrak X_c$ is the set of all edges connecting $\gamma_i$ and $\gamma_j$, and a set of edges $\{e_{ij}\}$, one from each non-empty set $\mathfrak X_c$ (sec.~\ref{sec:coarse-graining edges});
	
	\item a set of geometric tags $\{\vartheta_i,\varepsilon_i\}$, each associated to a subgraph $\gamma_i$ and constructed following the method described in sec.~\ref{sec:coarse-graining tags}.
\end{itemize}
These correspond to all the choices required to perform the coarse-graining prescriptions described earlier. We collectively denote these inputs $\rho$. For each admissible choice $\rho$, $\mathcal K_{\Gamma}^{(n,p)}$ maps the states in ${\tilde{\mathcal H}}_{\Gamma}$ to states supported on graphs with at most $n$ vertices, and each pair of vertices is connected by at most $p$ edges. Concretely, we define $\mathcal K_{\Gamma}^{(n,p)}$ through its action on spin network states in ${\tilde{\mathcal H}}_{\Gamma}$. Namely, given a normalized spin network state $\Psi_\Gamma$ and denoting by $\nu_\Gamma$ the number of vertices in $\Gamma$, we have:
\begin{enumerate}
	\itemsep0em
	
	\item if $\nu_\Gamma \leq n$ and the number of edges between any pair of vertices does not exceed $p$, then $\mathcal K_{\Gamma}^{(n,p)}$ acts simply as the identity map on $\Psi_\Gamma$: $\mathcal K_{\Gamma}^{(n,p)} \left[\rho, \Psi_\Gamma \right] = \Psi_\Gamma$;
	
	\item if $\nu_\Gamma \leq n$ and there are pairs of vertices which are connected by more than $p$ edges, then $\mathcal K_{\Gamma}^{(n,p)}$ acts by coarse-graining the sets of edges between such pairs of vertices, and the corresponding geometric tags. The resulting state is normalized\footnote{The normalization is realized using the inner product on ${\mathcal H}_{\rm T}$, since these states are also elements of ${\mathcal H}_{\rm T}$.} to obtain the image state of $\Psi_\Gamma$;
	
	\item if $\nu_\Gamma > n$, then $\mathcal K_{\Gamma}^{(n,p)}$ acts by first coarse-graining the vertices and geometric tags of $\Psi_\Gamma$ to obtain a state with $n$ vertices. Followed, if necessary, by a coarse-graining of the external edges and the corresponding geometric tags in the new state as in step $2.$ just above. Again, the resulting state is normalized to obtain the final image state of $\Psi_\Gamma$.
\end{enumerate}
For each graph $\Gamma$, there is a finite number of possible parameters $\rho$. Therefore, in order to eliminate this redundancy in the coarse-graining, we propose to use a straightforward averaging over the finite set of parameters $\mathcal R_\Gamma^{(n,p)}$. Namely, we define a new family of linear maps $\mathcal J_{\Gamma}^{(n,p)}$ from ${\tilde{\mathcal H}}_{\Gamma}$ to ${\mathcal H}_{\rm T}$:
\begin{align}
	\mathcal J_{\Gamma}^{(n,p)}[\ \cdot\ ] \equiv \frac{1}{\varrho_\Gamma^{(n,p)}} \sum_{\rho \in \mathcal R_\Gamma^{(n,p)}} \mathcal K_{\Gamma}^{(n,p)} \left[\rho, \cdot\ \right] \ ,
\end{align}
where $\varrho_\Gamma^{(n,p)}$ is the cardinality of the set $\mathcal R_\Gamma^{(n,p)}$.
Generically, each map $\mathcal J_{\Gamma}^{(n,p)}$ would map a spin network state in ${\tilde{\mathcal H}}_{\Gamma}$ to a linear combination of spin network states on different graphs, each with at most $n$ vertices and at most $p$ edges between any pair of vertices. Additionally, graphs with the same number of vertices and the same number of independent loops strictly differ by the number of coarse edges (capped at $p$), the distribution of self-loops among the vertices, and the geometric tags at each vertex.

Having the collection of maps $\mathcal J_{\Gamma}^{(n,p)}$ on each space ${\tilde{\mathcal H}}_{\Gamma}$, we can introduce a densely defined linear map $\mathcal I_{n,p}$ on ${\mathcal H}_{\rm T}$ for any fixed $n,\, p \in \mathbbm N^+ \cup \{\infty\}$ as
\begin{align}
	\mathcal I_{n,p} \equiv \underset{\Gamma}{\otimes}\, \mathcal J_{\Gamma}^{(n,p)}\ ,
	\label{Coarse-graining_maps}
\end{align}
where the direct product over graphs has the same range as the direct sum in \eqref{decomp_HT}. Note that when $n=p=\infty$, $\mathcal I_{n,p}=\mathbbm 1$ on ${\mathcal H}_{\rm T}$.

From the construction of the maps $K_{\Gamma}^{(n,p)}$ and $J_{\Gamma}^{(n,p)}$, a map $\mathcal I_{n,p}$ is bounded on its domain given by the span of the set of all spin network states in ${\mathcal H}_{\rm T}$. This implies that $\mathcal I_{n,p}$ is closable and its closure is bounded on all $\mathcal H_{\rm T}$. Hence, from here on, we only consider the closures of the maps $\mathcal I_{n,p}$ and we use the same notation $\mathcal I_{n,p}$ for the closed extensions. Additionally, the maps $\mathcal I_{n,p}$ are projections: $\mathcal I_{n,p}^2 = \mathcal I_{n,p}$. We will further discuss the properties of these maps in section \ref{sec:Hamiltonian_Renormalization}.

The image of ${\mathcal H}_{\rm T}$ by a map $\mathcal I_{n,p}$ corresponds to a subspace of ${\mathcal H}_{\rm T}$, which is spanned by spin networks supported on graphs having at most $n$ vertices, at most $p$ edges between any pair of vertices, and an arbitrary number of self-loops distributed among them. As noted at the end of section \ref{sec:coarse-graining edges}, the coarse-graining can generate coarse-edges with zero spin, i.e.~a group element in the trivial representation. As a consequence, the image of ${\mathcal H}_{\rm T}$ by a map $\mathcal I_{n,p}$ is not a subset of any specific ${\tilde{\mathcal H}}_{\Gamma}$, but rather a subspace of an infinite direct sum of Hilbert spaces ${\tilde{\mathcal H}}_{\Gamma}$ that is not equal to ${\mathcal H}_{\rm T}$. 

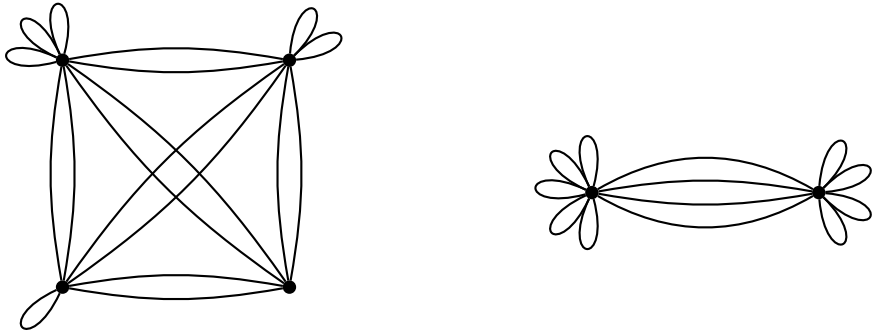
\begin{figure}[H]
	\centering
	\begin{tikzpicture}[thick,baseline=(v)]
		\node[circle,fill=black,draw=black,inner sep=1.5pt] (p0) at (0,0) {};
		\node[circle,fill=black,draw=black,inner sep=1.5pt] (p1) at (3,0) {};
		\node[circle,fill=black,draw=black,inner sep=1.5pt] (p2) at (3,3) {};
		\node[circle,fill=black,draw=black,inner sep=1.5pt] (p3) at (0,3) {};
		\draw[thick] (p0) to[out=10,in=170] (p1);
		\draw[thick] (p0) to[out=-10,in=-170] (p1);
		\draw[thick] (p0) to[out=55,in=-145] (p2);
		\draw[thick] (p0) to[out=35,in=-125] (p2);
		\draw[thick] (p0) to[out=100,in=-100] (p3);
		\draw[thick] (p0) to[out=80,in=-80] (p3);
		\draw[thick] (p1) to[out=100,in=-100] (p2);
		\draw[thick] (p1) to[out=80,in=-80] (p2);
		\draw[thick] (p3) to[out=10,in=170] (p2);
		\draw[thick] (p3) to[out=-10,in=-170] (p2);
		\draw[thick] (p1) to[out=145,in=-55] (p3);
		\draw[thick] (p1) to[out=125,in=-35] (p3);
		\draw[thick] (p0) to[out=-155,in=-115,looseness=40] (p0);
		\draw[thick] (p2) to[out=5,in=45,looseness=40] (p2);
		\draw[thick] (p2) to[out=45,in=85,looseness=40] (p2);
		\draw[thick] (p3) to[out=75,in=115,looseness=40] (p3);
		\draw[thick] (p3) to[out=115,in=155,looseness=40] (p3);
		\draw[thick] (p3) to[out=155,in=195,looseness=40] (p3);

		\node[circle,fill=black,draw=black,inner sep=1.5pt] (p4) at (10,1.25) {};
		\node[circle,fill=black,draw=black,inner sep=1.5pt] (p5) at (7,1.25) {};
		\draw[thick] (p5) to[out=10,in=170] (p4);
		\draw[thick] (p5) to[out=-10,in=-170] (p4);
		\draw[thick] (p5) to[out=30,in=150] (p4);
		\draw[thick] (p5) to[out=-30,in=-150] (p4);
		\draw[thick] (p4) to[out=5,in=45,looseness=40] (p4);
		\draw[thick] (p4) to[out=45,in=85,looseness=40] (p4);
		\draw[thick] (p4) to[out=-5,in=-45,looseness=40] (p4);
		\draw[thick] (p4) to[out=-45,in=-85,looseness=40] (p4);
		
		\draw[thick] (p5) to[out=75,in=115,looseness=40] (p5);
		\draw[thick] (p5) to[out=115,in=155,looseness=40] (p5);
		\draw[thick] (p5) to[out=155,in=195,looseness=40] (p5);
		\draw[thick] (p5) to[out=-75,in=-115,looseness=40] (p5);
		\draw[thick] (p5) to[out=-115,in=-155,looseness=40] (p5);
	\end{tikzpicture}
	\caption{Examples of multi-reflexive \mbox{$(n,p)$-complete} graphs: $(4,2)$-complete graph on the left, and $(2,4)$-complete graph on the right.}
	\label{fig:CG_example}
\end{figure}

In order to characterize the image of ${\mathcal H}_{\rm T}$ properly, we introduce the notion of multi-reflexive \mbox{$(n,p)$-complete} graph (see \figref{fig:CG_example} for an illustration):
\begin{quote}
	\emph{A multi-reflexive \mbox{$(n,p)$-complete} graph is a tagged abstract graph with exactly $n$ vertices, such that every vertex admits an arbitrary number of self-loops (reflexive edges), and any pair of vertices is connected by exactly $p$ edges.}
\end{quote}

We denote such graphs by $\Upsilon$, as opposed to $\Gamma$ for a generic graph, and we denote by $\mathcal Y_{(n,p)}$ the set of graphs $\Upsilon$ which have the same values of the parameters $n$ and $p$. The graphs in $\mathcal Y_{(n,p)}$ differ by the number of self-loops and sets of geometric tags at the $n$ vertices. Notice that in the case where $n=1$, the parameter $p$ becomes irrelevant and we will come back to this point once we define the image spaces.

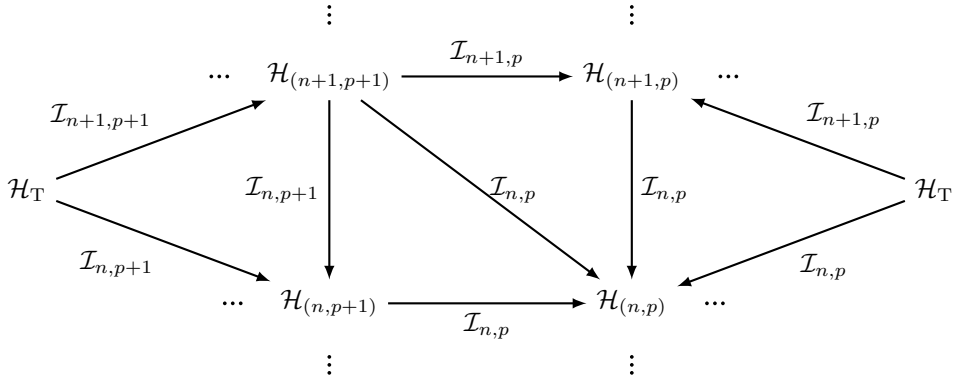
\begin{figure}[H]
	\centering
	\begin{tikzpicture}[thick]
		\node (H_npplus1) at (0,0) {$\mathcal{H}_{(n,p+1)}$};
		\node (H_nplus1pplus1) at (0,3) {$\mathcal{H}_{(n+1,p+1)}$};
		\node (H_np) at (4,0) {$\mathcal{H}_{(n,p)}$};
		\node (H_nplus1p) at (4,3) {$\mathcal{H}_{(n+1,p)}$};
		\node (H_T1) at (-4,1.5) {$\mathcal{H}_{\rm T}$};
		\node (H_T2) at (8,1.5) {$\mathcal{H}_{\rm T}$};
		\draw[-latex] (H_nplus1pplus1) -- node[midway,above] {$\mathcal I_{n+1,p}$} (H_nplus1p);
		\draw[-latex] (H_nplus1pplus1) -- node[midway,left] {$\mathcal I_{n,p+1}$} (H_npplus1);
		\draw[-latex] (H_nplus1pplus1) -- node[midway,right] {$\mathcal I_{n,p}$} (H_np);
		\draw[-latex] (H_npplus1) -- node[midway,below] {$\mathcal I_{n,p}$} (H_np);
		\draw[-latex] (H_nplus1p) -- node[midway,right] {$\mathcal I_{n,p}$} (H_np);
		\threedots{$(H_nplus1pplus1.north)+(0,0.5)$}[90]
		\threedots{$(H_nplus1pplus1.west)-(0.5,0)$}
		\threedots{$(H_nplus1p.north)+(0,0.5)$}[90]
		\threedots{$(H_nplus1p.east)+(0.5,0)$}
		\threedots{$(H_npplus1.south)-(0,0.5)$}[90]
		\threedots{$(H_npplus1.west)-(0.5,0)$}
		\threedots{$(H_np.south)-(0,0.5)$}[90]
		\threedots{$(H_np.east)+(0.5,0)$}
		\draw[-latex] (H_T1) -- node[midway,above left] {$\mathcal I_{n+1,p+1}$} (H_nplus1pplus1);
		\draw[-latex] (H_T1) -- node[midway,below left] {$\mathcal I_{n,p+1}$} (H_npplus1);
		\draw[-latex] (H_T2) -- node[midway,above right] {$\mathcal I_{n+1,p}$} (H_nplus1p);
		\draw[-latex] (H_T2) -- node[midway,below right] {$\mathcal I_{n,p}$} (H_np);
	\end{tikzpicture}
	\vspace*{1mm}
	\caption{Coarse-graining Hilbert spaces via the maps $\mathcal I_{n,p}$.}
	\label{CG_maps&HilbetSp.}
\end{figure}

From the discussion above, it follows that the image of ${\mathcal H}_{\rm T}$ by the projection $\mathcal I_{n,p}$ corresponds to a Hilbert space given by the completion of the span of spin network states supported on all the graphs in $\mathcal Y_{(n,p)}$, where the spins of the edges connecting any pair of vertices are allowed to vanish. We denote this Hilbert space $\mathcal H_{(n,p)}$ and we have: 
\begin{align}
	\mathcal H_{(n,p)} \equiv {\rm Ran}[\mathcal I_{n,p}]\ .
\end{align}
Taking into account the fact that the spin of a coarse edge can be zero, we introduce the sets $\{\tilde{\mathcal Y}_{(n,p)}\}$ where each $\tilde{\mathcal Y}_{(n,p)}$ consists of all the graphs in $\mathcal Y_{(n,p)}$ in addition to their subgraphs obtained by removing any number of edges connecting the vertices. This leads to expressing $\mathcal H_{(n,p)}$ as
\begin{align}
	\mathcal H_{(n,p)} = \bigoplus_{\Upsilon \in\, \bigcup\limits{_{k=0}^n}\tilde{\mathcal Y}_{(k,p)}} {\tilde{\mathcal H}}_{\Upsilon}\ ,
\end{align}
where the spaces ${\tilde{\mathcal H}}_{\Upsilon}$ are the ones in the decomposition of ${\mathcal H}_{\rm T}$ in \eqref{decomp_HT}.
Hence, starting from the Hilbert space ${\mathcal H}_{\rm T}$ we arrive at a family of \emph{coarse Hilbert spaces} $\mathcal H_{(n,p)}$, with $n,\, p \in \mathbbm{N}^+ \cup \infty$, corresponding to the images of ${\mathcal H}_{\rm T}$ by the linear maps $\mathcal I_{n,p}$ as depicted in \figref{CG_maps&HilbetSp.}.
The spaces $\mathcal H_{(n,p)}$ are subspaces of ${\mathcal H}_{\rm T}$, such that for any values $n,\, p \in \mathbbm{N}^+ \cup \infty$ and $m,\, q\in \mathbbm{N}$ they satisfy
\begin{align}\label{Hnp_nesting}
	\mathcal H_{(n,p)} \subset \mathcal H_{(n+m,p+q)}\ , \qquad \mathcal I_{n,p}[\mathcal H_{(n+m,p+q)}] = \mathcal H_{(n,p)}\ .
\end{align}
We then define the operator algebra $\mathfrak U_{n,p}$ on a Hilbert space $\mathcal H_{(n,p)}$ as the sub-algebra of the operator algebra $\mathfrak U$ on $\mathcal{H}_{\text{\rm T}}$ which preserves the space $\mathcal H_{(n,p)}$. Consequently, the admissible holonomy operators on a Hilbert space $\mathcal H_{(n,p)}$ are those supported by the multi-reflexive \mbox{$(n,p)$-complete} graph in $\mathcal Y_{n,p}$.

In summary, the \emph{coarse-graining maps} $\mathcal I_{n,p}$ provide a family of \emph{coarse Hilbert spaces} $\mathcal H_{(n,p)}$, where the labels $n$ and $p$ represent \emph{resolution} parameters or \emph{scales} of the coarse-graining process. The states in each space $\mathcal H_{(n,p)}$ are supported on a multi-reflexive \mbox{$(n,p)$-complete} graph.

This completes the kinematical structure of our effective theories: For each pair $(n,p) \in (\mathbbm N^+ \cup \{\infty\})^2$, which respectively correspond to the maximal number of vertices (nodes) and edges (paths) between any pair of vertices in a graph, we have a coarse Hilbert space $\mathcal H_{(n,p)}$, obtained via the coarse-graining map $\mathcal I_{n,p}$, and the associated operator algebra $\mathfrak U_{n,p}$. The three cases where one or both parameters $n$ and $p$ are equal to infinity are to be understood as follows: (i) the case where $n$ is finite and $p=\infty$ corresponds to the Hilbert space obtained by coarse-graining the vertices only, without coarse-graining the external edges; (ii) the case where $p$ is finite and $n=\infty$ corresponds to the Hilbert space obtained by coarse-graining the edges only, without coarse-graining the vertices; (iii) the case where $n=p=\infty$ corresponds simply to the initial Hilbert space $\mathcal{H}_{\text{\rm T}}$ where no coarse-graining occurs. Each of these Hilbert spaces is admissible in our construction of effective theories and coarse-grained models. Finally, as noted earlier, in the case where $n=1$ there are no edges connecting vertices as there is only one vertex. In fact, one can see from the construction of the maps $\mathcal I_{n,p}$ that if $n=1$, then $\mathcal I_{1,p} = \mathcal I_{1,q}$ and consequently $\mathcal H_{(1,p)} = \mathcal H_{(1,q)}$ for any $p$ and $q$ in $\mathbbm N^+ \cup \{\infty\}$. The space $\mathcal H_{(1,p)}$ is supported on a single-vertex graph with an arbitrary number of self-loops, or the $(1,0)$-complete graph in the terminology introduced earlier; we nonetheless retain the general label $\mathcal H_{(1,p)}$, rather than switching to $\mathcal H_{(1,0)}$, for notational uniformity across the family of coarse Hilbert spaces.

The next step in the definition of the effective theories is to construct an effective Hamiltonian operator on the coarse Hilbert spaces. 
Our approach, presented in the following section, relies on identifying and classifying effective interactions obtained from the actions of the Hamiltonian operators defined in the fundamental theory. As we will see, different regularizations induce different types of interaction among the vertices of the \mbox{$(n,p)$-complete} graphs.


\section{Dynamics for coarse states and effective theories}\label{sec:Effective Dynamics}

Before turning to the construction of the effective dynamics for the coarse states in the spaces $\mathcal H_{(n,p)}$, we would like to explain the perspective and the approach we follow. 

First, as mentioned in the Introduction, we treat the dynamics in the context of a deparametrized setting, where the Hamiltonian generates evolution of states and observables via a Schrödinger-like equation. The primary reason for not considering the fully constrained theory is that the physical states, i.e.~solutions to the Hamiltonian constraint equation, are not necessarily elements of the kinematical Hilbert spaces. Indeed, the consensus is that the physically relevant solutions are not normalizable with respect to the kinematical inner product, and the full space of solutions would require a different physical inner product. This fact is likely to affect the coarse-graining procedures we presented above, as well as the construction of the effective dynamics and the corresponding renormalization flow equations. In contrast to the fully constrained theory, the treatment of the dynamics in the deparametrized setting is conceptually clear and the kinematical structures we defined so far represent the physical sectors.

Second, in our approach to constructing an effective Hamiltonian, we rely on the identification of the various types of actions induced by the different Hamiltonian operators available in the fundamental theory, i.e.~canonical LQG, on the coarse states. This provides an explicit connection between the dynamics in the fundamental theory and the effective dynamics at the coarse-grained level. Furthermore, given that we also propose to use the Hamiltonian renormalization framework to control the effective dynamics when changing the coarse-graining scales, this approach enables treating the fundamental dynamics in an agnostic manner. Namely, we consider all the different regularizations proposed so far in the literature for the Hamiltonian constraint operator, then treat the various prescriptions as ambiguities which could be reduced (if not removed) through coarse-graining and renormalization. 

Operator ordering and symmetrizations aside, the regularizations of the gravitational Hamiltonian or Yang-Mills Hamiltonian primarily differ in the prescriptions of the loops associated to the holonomy operator, which replaces the classical connection and its curvature. These prescriptions can be split into two categories: The graph-preserving regularizations and the graph-changing ones. In the following, (i) we briefly overview the quantization of the classical Hamiltonian functional for gravity, and classify the various terms involved via the action of the corresponding operators; (ii) we identify the effective action of each term on the coarse states and define the types of interaction on the coarse vertices; (iii) we provide our proposal for the form of the effective Hamiltonian on a given coarse Hilbert space $\mathcal H_{(n,p)}$, as well as the definition of our effective theories.


\subsection{Hamiltonian operators in loop quantum gravity}\label{sec:Hamil_Op}

In the Ashtekar-Barbero formulation, the classical Hamiltonian constraint functional for gravity with a cosmological constant $\Lambda$ takes the form
\begin{equation}\label{Hamil_Const}
	\begin{aligned}
	C[N] &= -\frac{1}{2\kappa\beta^2}\int_\Sigma\,d^3x\, N \qty[\frac{\epsilon\indices{^{ij}_k}E^a_iE^b_jF^k_{ab}}{\sqrt{|\operatorname{det}(E)|}} + (1+\beta^2)\sqrt{|\operatorname{det}(E)|}R[E] - 2\beta^2\Lambda\sqrt{|\operatorname{det}(E)|}] \\[1em]
	&= \frac{1}{2\kappa}\int_\Sigma\,d^3x\, N \qty[\frac{\epsilon\indices{^{ij}_k}E^a_iE^b_j}{\sqrt{|\operatorname{det}(E)|}} \qty(F^k_{ab}-(1+\beta^2) \epsilon\indices{^k_{lm}} K_a^l K_b^m) + 2\Lambda\sqrt{|\operatorname{det}(E)|}]\ .
	\end{aligned}
\end{equation}
where $N$ is the lapse function, $F \equiv dA+A\wedge A$ is the curvature 2-form of the connection $A$, $R[E]$ is the three-dimensional Ricci scalar and $K_a$ is the extrinsic curvature 1-form.
These expressions are the starting point for the definition of a Hamiltonian operator in canonical LQG. In the deparametrized context, the physical Hamiltonian is a function $H$ of the functional $C[N]$ above. For instance, in the Gaussian dust model \cite{Kuchar:1990vy} $H = -C[1]$, while in the model with a minimally coupled massless scalar field \cite{Domagala:2010bm} we have that $H \propto C\qty[|\operatorname{det}(E)|^{1/2}]{}^{1/2}$. The lapse is actually determined by requiring the conservation of the chosen time gauge. Consequently, the quantization of the physical Hamiltonian involves the same prescriptions used to quantize the Hamiltonian constraint. In order to keep the discussion simple in what follows, we assume that the lapse is a spacetime function and not a phase space functional. The latter case can be treated with the same regularization prescriptions, upon determination of the induced classical Hamiltonian functional.

The quantization of the functional $C$ begins by regularizing the terms in \eqref{Hamil_Const}, i.e.~writing them in terms of holonomies and fluxes. We denote these different terms as
\begin{equation}\label{Hamil_terms}
	\begin{aligned}
		C[N] &= -\frac{1}{2\kappa\beta^2} \qty[ H_E[N] + (1+\beta^2) H_R[N] - 2 \beta^2 \Lambda V[N] ] \\
		&= \frac{1}{2\kappa} \qty[ H_E[N] - (1+\beta^2) H_L[N] + 2 \Lambda V[N] ]\ ,
	\end{aligned}
\end{equation}
and we can also write
\begin{equation}\label{Combined_Hamil_terms}
	\begin{aligned}
		C[N] &= \frac{1}{4\kappa\beta^2} \qty[ (\beta^2 - 1) H_E[N] - (1+\beta^2) H_R[N] - \beta^2(1+\beta^2) H_L[N] + 4 \beta^2 \Lambda V[N] ]\ ,
	\end{aligned}
\end{equation}
where $H_E$ is the so-called Euclidean constraint given by the term which contains the curvature $F$, $H_R$ is the densitized Ricci scalar term, $V$ is the smeared spatial volume, and $H_L$ is the extrinsic curvature term often called the Lorentzian part.
Each term in \eqref{Hamil_terms} is regularized separately with respect to a graph of a spin network state and the general idea is to map the densitized triads $E$ to fluxes, and the connection and its curvature to holonomies along open edges and closed loops respectively. Some terms such as the Ricci scalar and the extrinsic curvature 1-form $K_a$ require intermediate steps to arrive at \emph{quantizable} expressions. Namely, the Ricci term is discretized using identities from Regge calculus \cite{Alesci:2014aza}, while $K_a$ is written in terms of Poisson brackets \cite{Thiemann:1996au} involving the connection $A$, the spatial volume $V[1]$ and the Euclidean term $H_E[1]$. Once the regularized expressions are derived, one simply maps the classical holonomies and fluxes to their quantum counterparts for a given graph, then one takes the limit of the regulator going to zero to obtain the final operators. In appendix \ref{Expressions_Hamiltonian}, we provide the expressions of the operators corresponding to each term in \eqref{Hamil_Const}, denoted with extra hats with respect to their classical counterparts.

Most of the details of the regularization procedures are not relevant for our construction of the effective dynamics; full accounts can be found in  \cite{Rovelli:1993bm, Rovelli:1994ge, Ashtekar:1997fb, Thiemann:1996aw, Thiemann:2003zv, Giesel:2006uj, Alesci:2014aza, Alesci:2015wla, Assanioussi:2015gka}. What matters for us are the general features of the actions of the quantum operators associated to each term in $C$, and how they alter the data of a spin network state. In fact, in our approach which we develop in the next section, the form of each term involved in the effective Hamiltonian does not differ from the expression of its counterpart in the fundamental Hamiltonian operator. In other words, the terms in the effective Hamiltonian retain the exact same dependence on the holonomy and flux operators (or equivalently the $J$ operators). The difference resides in the graph structure associated to the holonomies involved. The reason for this requirement on the form of the terms involved in the effective Hamiltonian stems from the expectation of matching the classical expressions in a semi-classical regime of an effective theory or coarse-grained model. In the absence of a clear and justified prescription, a modification of the dependence on the holonomy and flux operators could make the recovery of the semi-classical limit, even for geometric observables such as the spatial volume, dramatically more complicated. Since the semi-classical limit is one of the ultimate tests of a quantum model, retaining the same operator forms for the Hamiltonian at the effective level seems to be an advantage at this stage.

As mentioned earlier, the primary difference between the various regularizations is whether the Hamiltonian operator is graph-preserving or not. The $J$ operators do not change the graphs; hence, from the expressions in \eqref{V_AL} and \eqref{H_R} we have that the volume $\hat V$ and Ricci curvature $\hat H_R$ operators are graph-preserving. Also, from the expression of the Lorentzian part $\hat H_L$ in \eqref{H_L}, it follows that whether $\hat H_L$ preserves the graph or not depends solely on whether the Euclidean part $\hat H_E$ does or doesn't.
Therefore, the aspect of whether the Hamiltonian operator preserves the graph or not is exclusively encoded in the regularization of the Euclidean term $\hat H_E$, and specifically in the holonomy around the closed loop which replaces the curvature $F$. There are three possible prescriptions for such loops: (i) the loop overlaps with the pre-existing graph, which is the graph-preserving prescription \cite{Giesel:2006uj}; (ii) the loop partially overlaps with the pre-existing graph, which is a graph-changing prescription rigorously implemented as the \emph{special edge} regularization introduced in \cite{Thiemann:1996aw}; (iii) the loop is a self-loop and does not overlap with the pre-existing graph, which is also a graph-changing prescription rigorously implemented as the \emph{special loop} regularization introduced in \cite{Alesci:2015wla, Assanioussi:2015gka}. Finally, while it is obvious that all the operators in the Hamiltonian change the intertwiner states they act on, not all terms affect the spins associated to the edges of the graph of the state. In fact, only the Euclidean part $\hat H_E$, and consequently the Lorentzian term $\hat H_L$, can generate spin coupling, and it is again only through the holonomies around the closed loops when these are chosen to overlap with the pre-existing graph (i.e.~in the prescriptions (i) and (ii) above). We summarize all these properties in Table~\ref{tab:Hamil_Operators}.
\begin{table}[t!]
	\centering
	\caption{Properties induced by the action of the different terms in the Hamiltonian operator.}
	\label{tab:Hamil_Operators}
	\addtolength{\tabcolsep}{1em}
	\begin{tabular}{cccc}
		\toprule
		\textbf{Operators} & \textbf{Loop holonomy} & \textbf{Spin coupling} & \textbf{Graph-changing} \\
		\midrule
		$\hat V$ & No & No & No  \\
		$\hat H_R$ & No & No & No \\
		$\hat H_E$ and $\hat H_L$ & Yes, complete overlap & Yes & No \\
		$\hat H_E$ and $\hat H_L$ & Yes, partial overlap & Yes & Yes \\
		$\hat H_E$ and $\hat H_L$ & Yes, no overlap & No & Yes \\
		\bottomrule
	\end{tabular}
	\addtolength{\tabcolsep}{-1em}
\end{table}

We can now proceed to the construction of the effective Hamiltonian for the coarse states by identifying the various \emph{admissible interaction} terms.


\subsection{Vertex-interactions and the effective Hamiltonian}\label{sec:Vertex_interactions}

As a consequence of the coarse-graining procedure, the maps $\mathcal I_{n,p}$ preserve the number of independent loops in gauge invariant spin network states.
Given that the graphs of the coarse states are supported on multi-reflexive \mbox{$(n,p)$-complete} graphs, a coarse-graining map $\mathcal I_{n,p}$ would map a spin network state and its image by an operator to the same Hilbert space $\mathcal H_{(n,p)}$, but to states with possibly different numbers of self-loops at the vertices. Therefore, a graph-preserving operator in the fundamental theory, i.e.~not changing the number of independent loops, is to be represented by a graph-preserving operator on the coarse Hilbert spaces; and a graph-changing operator which creates or destroys loops is to be represented by an operator which respectively adds or removes self-loops at the coarse vertices of the \mbox{$(n,p)$-complete} graphs. In particular, self-loops remain self-loops after applying the coarse-graining procedure presented above.

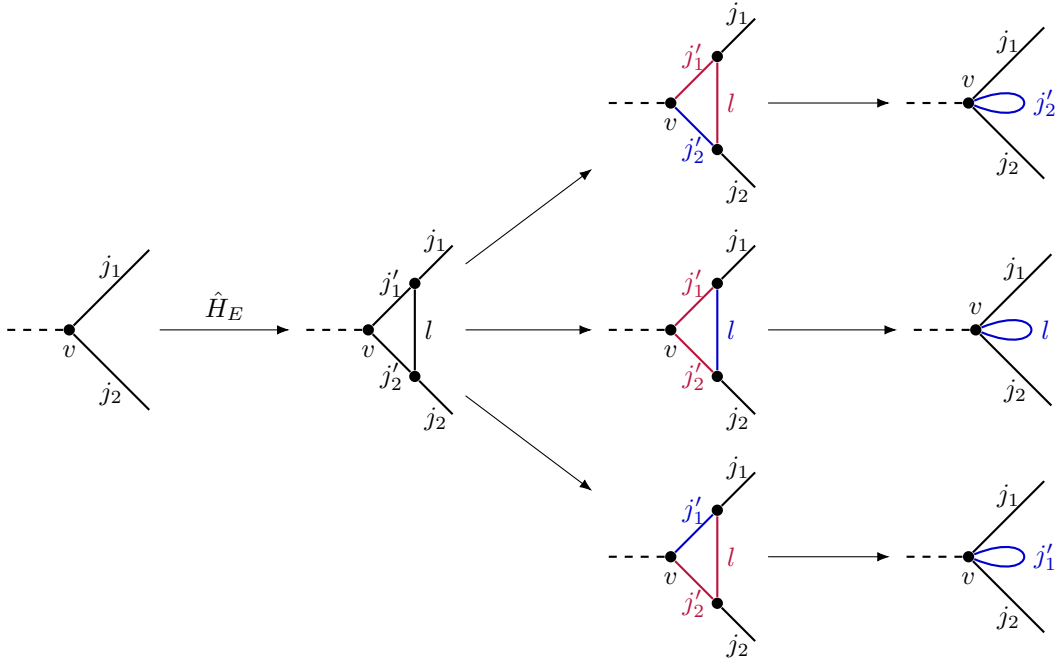
\begin{figure}[!t]
	\centering
	\begin{tikzpicture}[baseline=(ini),every edge/.style={very thick}]
		\node (ini) at (0,0) {
			\begin{tikzpicture}[baseline=(p0)]
				\node[p_node,label=below:{$v$}] (p0) at (0,0) {};
				\coordinate[above right=of p0] (p1) {};
				\coordinate[below right=of p0] (p2) {};
				\draw[thick,dashed] (p0) -- ++(-0.9,0);
				\draw[thick] (p0) --node[midway,above]{$j_1$} (p1);
				\draw[thick] (p0) --node[midway,below]{$j_2$} (p2);
			\end{tikzpicture}
		};
		\node (H_E action) at (4,0) {
			\begin{tikzpicture}[baseline=(p0)]
				\node[p_node,label=below:{$v$}] (p0) at (0,0) {};
				\node[p_node,above right=0.5 and 0.5 of p0] (p1) {};
				\node[p_node,below right=0.5 and 0.5 of p0] (p2) {};
				\draw[thick,dashed] (p0) -- ++(-0.9,0);
				\draw[thick] (p0) --node[midway,above]{$j^\prime_1$} (p1);
				\draw[thick] (p0) --node[midway,below]{$j^\prime_2$} (p2);
				\draw[thick] (p1) --node[midway,right]{$l$} (p2);
				\draw[thick] (p1) --node[midway,above]{$j_1$} ++(0.5,0.5);
				\draw[thick] (p2) --node[midway,below]{$j_2$} ++(0.5,-0.5);
			\end{tikzpicture}
		};
		\node (triv 1) at (8,3) {
			\begin{tikzpicture}[baseline=(p0)]
				\node[p_node,label=below:{$v$}] (p0) at (0,0) {};
				\node[p_node,above right=0.5 and 0.5 of p0] (p1) {};
				\node[p_node,below right=0.5 and 0.5 of p0] (p2) {};
				\draw[thick,dashed] (p0) -- ++(-0.9,0);
				\draw[thick,NCBJred] (p0) --node[midway,above]{$j^\prime_1$} (p1);
				\draw[thick, blue!80!black] (p0) --node[midway,below]{$j^\prime_2$} (p2);
				\draw[thick,NCBJred] (p1) --node[midway,right]{$l$} (p2);
				\draw[thick] (p1) --node[midway,above]{$j_1$} ++(0.5,0.5);
				\draw[thick] (p2) --node[midway,below]{$j_2$} ++(0.5,-0.5);
			\end{tikzpicture}
		};
		\node (triv 2) at (8,-3) {
			\begin{tikzpicture}[baseline=(p0)]
				\node[p_node,label=below:{$v$}] (p0) at (0,0) {};
				\node[p_node,above right=0.5 and 0.5 of p0] (p1) {};
				\node[p_node,below right=0.5 and 0.5 of p0] (p2) {};
				\draw[thick,dashed] (p0) -- ++(-0.9,0);
				\draw[thick, blue!80!black] (p0) --node[midway,above]{$j^\prime_1$} (p1);
				\draw[thick,NCBJred] (p0) --node[midway,below]{$j^\prime_2$} (p2);
				\draw[thick,NCBJred] (p1) --node[midway,right]{$l$} (p2);
				\draw[thick] (p1) --node[midway,above]{$j_1$} ++(0.5,0.5);
				\draw[thick] (p2) --node[midway,below]{$j_2$} ++(0.5,-0.5);
			\end{tikzpicture}
		};
		\node (triv 3) at (8,0) {
			\begin{tikzpicture}[baseline=(p0)]
				\node[p_node,label=below:{$v$}] (p0) at (0,0) {};
				\node[p_node,above right=0.5 and 0.5 of p0] (p1) {};
				\node[p_node,below right=0.5 and 0.5 of p0] (p2) {};
				\draw[thick,dashed] (p0) -- ++(-0.9,0);
				\draw[thick,NCBJred] (p0) --node[midway,above]{$j^\prime_1$} (p1);
				\draw[thick,NCBJred] (p0) --node[midway,below]{$j^\prime_2$} (p2);
				\draw[thick, blue!80!black] (p1) --node[midway,right]{$l$} (p2);
				\draw[thick] (p1) --node[midway,above]{$j_1$} ++(0.5,0.5);
				\draw[thick] (p2) --node[midway,below]{$j_2$} ++(0.5,-0.5);
			\end{tikzpicture}
		};
		\node (coll 1) at (12,3) {
			\begin{tikzpicture}[baseline=(p0)]
				\node[p_node,label=above:{$v$}] (p0) at (0,0) {};
				\draw[thick,dashed] (p0) -- ++(-0.9,0);
				\draw[thick] (p0) --node[midway,above]{$j_1$} ++(1,1);
				\draw[thick] (p0) --node[midway,below]{$j_2$} ++(1,-1);
				\draw[thick,blue!80!black] (p0) to[in=-22.5,out=22.5,looseness=40] node[midway,right]{$j^\prime_2$} (p0);
			\end{tikzpicture}
		};
		\node (coll 2) at (12,-3) {
			\begin{tikzpicture}[baseline=(p0)]
				\node[p_node,label=below:{$v$}] (p0) at (0,0) {};
				\draw[thick,dashed] (p0) -- ++(-0.9,0);
				\draw[thick] (p0) --node[midway,above]{$j_1$} ++(1,1);
				\draw[thick] (p0) --node[midway,below]{$j_2$} ++(1,-1);
				\draw[thick,blue!80!black] (p0) to[in=-22.5,out=22.5,looseness=40] node[midway,right]{$j^\prime_1$} (p0);
			\end{tikzpicture}
		};
		\node (coll 3) at (12,0) {
			\begin{tikzpicture}[baseline=(p0)]
				\node[p_node,label=above:{$v$}] (p0) at (0,0) {};
				\draw[thick,dashed] (p0) -- ++(-0.9,0);
				\draw[thick] (p0) --node[midway,above]{$j_1$} ++(1,1);
				\draw[thick] (p0) --node[midway,below]{$j_2$} ++(1,-1);
				\draw[thick,blue!80!black] (p0) to[in=-22.5,out=22.5,looseness=40] node[midway,right]{$l$} (p0);
			\end{tikzpicture}
		};
		\draw[-Latex] (ini) --node[midway,above]{$\hat{H}_E$} (H_E action);
		\draw[-Latex] (H_E action) -- (triv 1);
		\draw[-Latex] (H_E action) -- (triv 2);
		\draw[-Latex] (H_E action) -- (triv 3);
		\draw[-Latex] (triv 1) --node[midway,above]{} (coll 1);
		\draw[-Latex] (triv 2) --node[midway,above]{} (coll 2);
		\draw[-Latex] (triv 3) --node[midway,above]{} (coll 3);
	\end{tikzpicture}
	\caption{An example illustrating the coarse-graining of the action of the operator $\hat H_E$, adding a special edge, with respect to different choice of trees (in red).}
	\label{CG_special_edge}
\end{figure}

Using these observations, combined with the requirement of retaining the same dependence on holonomies and fluxes, we can deduce the various effective terms induced by the operators $\hat V$, $\hat H_R$, $\hat H_E$ and $\hat H_L$ involved in the Hamiltonians of the fundamental theory. Therefore, on a given coarse Hilbert space $\mathcal H_{(n,p)}$ we have:
\begin{itemize}
	\item The effective volume operator $\check V$ is given by the restriction of the original volume operator to the coarse space $\mathcal H_{(n,p)}$:
	\begin{align}
		\check V \equiv \hat V |_{\mathcal H_{(n,p)}}\ .
	\end{align}
	
	\item The effective Ricci operator $\check H_R$ is given by the restriction of the original Ricci operator to the coarse space $\mathcal H_{(n,p)}$:
	\begin{align}
		\check H_R \equiv \hat H_R |_{\mathcal H_{(n,p)}}\ .
	\end{align}
	
	\item The effective Euclidean operator has the same form as $\hat H_E$, with the use of $\check V$ as volume operator, and we have three types of actions:
	\begin{enumerate}[(i)]
		\item graph-preserving action: we have an effective operator $\check H_E^o$ which acts in a graph-preserving fashion and generates spin couplings around all the loops in a given graph, including the self-loops. At every vertex, each pair of edges selects a specific set of loops. Namely, if the pair of edges forms a self-loop at a vertex, then the set of admissible loops for the holonomy in the expression of $\check H_E^o$ consists of that single self-loop. The same applies if the pair of edges forms a loop between exactly two vertices, that is they have the same source and target vertices.
		However, if the pair of edges does not form a self-loop and does not connect the same two vertices, then the set of admissible loops consists of all the loops containing this pair of edges, without going through any vertex more than once (which automatically excludes self-loops in them). For an \mbox{$(n,p)$-complete} graph, the set of admissible loops is finite and the loops can be characterized by the number of vertices they go through. Specifically, for any pair of edges at a vertex, the number of loops they belong to and containing $m \geq 3$ vertices is simply $p^{m-2}$. In the context of the fundamental LQG theory, the graph-preserving prescription imposes that the loops are minimal loops, i.e.~formed by the smallest number of edges, with an extra averaging when there is more than one minimal loop for a given pair. In principle, one could adopt the same restriction for the effective operator, which would make the admissible loops be self-loops and loops formed by two or three coarse edges. However, such a restriction seems too strong from the perspective of the coarse-graining procedure, since minimal loops in the original states are not necessarily minimal loops at the coarse level and vice-versa. Furthermore, as we are planning to impose a renormalization flow equation on the effective Hamiltonian, a better alternative is to consider all the terms with all admissible loops, and introduce different coupling coefficients which would be determined by the flow equation rather than fixing them by hand. Therefore, on a coarse Hilbert space $\mathcal H_{(n,p)}$, the expression of the effective Euclidean operator $\check H_E^o$ is given as
		\begin{align}
			\check H_E^o \equiv \sum_{v} \left[ \sum_{\alpha_{IJ}^{(1)}} \hat H_E[\alpha_{IJ}^{(1)}] + \sum_{\alpha_{IJ}^{(2)}} \hat H_E[\alpha_{IJ}^{(2)}] + \frac{p-1}{p(p^{\nu-2}-1)}\sum_{m=3}^\nu \sum_{\alpha_{IJ}^{(m)}} \hat H_E[\alpha_{IJ}^{(m)}] \right]\ ,
		\label{HEo}
		\end{align}
		where the first sum is over the vertices and the other sums are over sets of existing loops, such that $\nu$ is the number of vertices in the state and $\alpha_{IJ}^{(m)}$ denotes a loop going through $m$ vertices. In particular, $\alpha_{IJ}^{(1)}$ corresponds to the self-loops and $\alpha_{IJ}^{(2)}$ corresponds to the single loop connecting a pair of vertices and containing the edges $I,J$. The factor ${p-1}/{p(p^{\nu-2}-1)}$ is the averaging factor over the number of loops present in the sum. The operators $\hat H_E[\alpha_{IJ}^{(m)}]$ are given in \eqref{H_E[IJ]} and \eqref{C_E[IJ]}, and we denote their sum over the loops with the same number of vertices by $\check H_E^{o(m)}$.
		
		\item graph-changing action with a special loop: we have an effective operator $\check H_E^+$ which adds and removes self-loops at the coarse vertices with a fixed spin $l$. The expression of $\check H_E^+$ is simply given by the restriction of the operator $\hat H_E$ to the space $\mathcal H_{(n,p)}$:
		\begin{align}
			\check H_E^+ \equiv \hat H_E |_{\mathcal H_{(n,p)}}\ .
		\end{align}
		In the original prescription \cite{Alesci:2015wla, Assanioussi:2015gka} to define the operator $\hat H_E^+$, the special loops are self-loops which satisfy certain geometric conditions at the vertices. Namely, a special loop is associated to a pair of edges at the vertex, and its edges are tangent to this pair of edges. This implies that at the level of states on tagged abstract graphs, the tags involving the edges of the special loop are directly obtained from the tags of the associated pair of edges. This property is systematically passed to the effective operator $\check H_E^+$.
		
		\item graph-changing action with a special edge: In this case we have an effective operator $\check H_E^-$ which also adds and removes self-loops at the coarse vertices. The prescription for the special edge regularization developed in \cite{Thiemann:1996aw} ensures that the edges of the self-loop are tangent to the associated pair of edges, leading to adopting the same tangentiality condition as for the self-loop created and destroyed by $\check H_E^+$. However, the difference with respect to $\check H_E^+$ lies in the spin associated to the self-loop. To determine the spin, we look at what happens if we coarse-grain a subgraph where the operator $\hat H_E[\alpha_{IJ}]$ has acted on a vertex $v$, as we illustrate in \figref{CG_special_edge}. The action of $\hat H_E[\alpha_{IJ}]$ creates (or destroys) a loop at $v$, consisting of two segments and a special edge connecting two three-valent vertices. The two segments are the overlap with the pre-existing graph and the associated spins are obtained from the coupling of the original spins with the spin of the loop $l$. Say that before the action of $\hat H_E[\alpha_{IJ}]$, we chose a tree to coarse-grain the initial state. This tree would either contain both segments, one of them only, or none of them. Assuming we use the same tree to coarse-grain the image state by $\hat H_E[\alpha_{IJ}]$, then we have three cases: If the tree contains the two segments, then the coarse-graining of the image would generate a coarse-vertex with an additional self-loop with spin $l$ with respect to the coarse-vertex obtained from the initial state; if the tree does not contain one of the segments or both, then the coarse-graining procedure requires completing the tree to a maximal tree passing through the three-valent vertices of the loop. 
		In the case where the tree already contains one segment, then one has to add either the second segment or the special edge to the tree. While in the case where none of the segments is part of the tree, one has to add both segments or one of them with the special edge. In either case, the additional self-loop after coarse-graining would have a spin $l$ or a spin provided by the coupling of the original spins with the spin $l$. It follows that the effective operator $\check H_E^-$ must be given by a sum of contributions which only differ by the spin they associate to the self-loop, namely
		\begin{align}
			\check H_E^- \equiv \sum_v \sum_{I,J} \sum_{s\in \mathfrak s_{IJ}} \hat H_E^{(s)}[\alpha_{IJ}] \ ,
		\end{align}
		where the first two sums are as in the definitions in the appendix, while the third sum is over the spin $s$ associated to the self-loop. The set of spin values is given by $\mathfrak s_{IJ} \equiv \{l\} \cup \llbracket \abs{j_I-l}, j_I+l \rrbracket \cup \llbracket \abs{j_J-l}, j_J+l \rrbracket$, with $j_I$ and $j_J$ being the spins of the edges $I$ and $J$ respectively and the brackets $\llbracket\ ,\ \rrbracket$ denote a closed and discrete interval where values vary by a step of one. Note that the term $\hat H_E^{(l)}[\alpha_{IJ}]$ coincides with the operator $\check H_E^+$ above.
		
	\end{enumerate}
	
	\item The effective Lorentzian part operator $\check H_L$ is straightforwardly obtained by replacing the volume operator $\hat V$ and Euclidean operator $\hat H_E$ by their chosen effective counterparts in the expression of the original operator $\hat H_L$. We then have three operators: $\check H_L^o$, $\check H_L^+$, $\check H_L^-$ depending quadratically on $\check H_E^o$, $\check H_E^+$ and $\check H_E^-$ respectively. 
	$\check H_L^o$ would preserve the graph, while the operators $\check H_L^+$ and $\check H_L^-$ would add and remove two self-loops at each vertex.

\end{itemize}

By construction, all the effective operators defined above preserve the Hilbert spaces $\mathcal H_{(n,p)}$ separately. Furthermore, on a single space $\mathcal H_{(n,p)}$, their actions can be viewed as dynamical interactions among the vertices of the graphs. Consequently, we are able to classify the effective operators with respect to the number of coarse vertices that their actions involve, in a fashion that mimics Feynman diagrams in standard quantum field theory. In fact, such a classification of interactions is not unusual in the context of quantum gravity, and a strong parallel can be made between the structure of the effective dynamics presented here and the perturbative approach used in \emph{group field theories} \cite{Oriti:2006se, Oriti:2013aqa}. Table~\ref{tab:Classification_Dyn_Int} summarizes the classification of the various effective dynamical terms introduced above.
\begin{table}[H]
	\caption{Classification of the effective operators by types of the induced dynamical interaction.}
	\label{tab:Classification_Dyn_Int}
	\addtolength{\tabcolsep}{1em}
	\centering
	\begin{tabular}{ccc}
		\toprule
		\textbf{Interaction type} & \textbf{Graph-preserving} & \textbf{Graph-changing} \\
		\midrule
		$1$-vertex & $\check V$, $\check H_R$, $\check H_E^{o(1)}$, $\check H_L^o$ & $\check H_E^+$, $\check H_E^-$, $\check H_L^+$, $\check H_L^-$  \\[1em]
		$m$-vertex ($m\geq2$) & $\check H_E^{o(m)}$, $\check H_L^o$ & - \\
		\bottomrule
	\end{tabular}
	\addtolength{\tabcolsep}{-1em}
\end{table}

The \emph{$1$-vertex} interactions consist of all mappings which involve the data at a single vertex only. These mappings can be graph-preserving by acting only in the intertwiner space (this includes all operators depending exclusively on the fluxes such as the volume and Ricci operators), or by changing the spins of the self-loops at the vertices. They can also be graph-changing by adding and removing self-loops. Intuitively, such mappings represent self-interactions at the vertex.

The \emph{$m$-vertex} or \emph{many-vertex} interactions consist of all mappings which involve many vertices at once. Such mappings \emph{couple} the vertices of the graph and can be interpreted as interaction terms among many vertices. The simplest example is the holonomy around a loop containing several vertices, present in the graph-preserving version $\check H_E^o$ of the Euclidean operator and by construction also in $\check H_L^o$.

This concludes the identification and classification of the effective interaction terms arising from the fundamental Hamiltonian dynamics of canonical loop quantum gravity. We may now introduce our proposal of the effective Hamiltonian for the coarse states on \mbox{$(n,p)$-complete} graphs.

As we mentioned earlier, our idea to construct the effective Hamiltonian is to consider all possible interaction terms induced by the quantization of the classical Hamiltonian functional, with relative coupling coefficients. These coefficients will become the parameters in the renormalization flow equations. Therefore, based on the above analysis of the different prescriptions and operators for the fundamental dynamics, we define an effective Hamiltonian on a given Hilbert space $\mathcal H_{(n,p)}$ as
\begin{align}
	\check H_n[\{\lambda_c\}] \equiv \lambda_V \check V + \lambda_R \check H_R + \lambda_E^+ \check H_E^+ + \lambda_E^- \check H_E^- + \sum\limits_{i=1}^n \lambda_E^{(i)} \check H_E^{o(i)} + \lambda_L^+ \check H_L^+ + \lambda_L^- \check H_L^- + \lambda_L^o \check H_L^o \ ,
	\label{Effective_Hamiltonian}
\end{align}
where $\{\lambda_c\}$ are arbitrary coupling coefficients. This Hamiltonian preserves the Hilbert space $\mathcal H_{(n,p)}$ and annihilates the vacuum state $\Omega_{\rm AL}$. This concludes the construction of the effective dynamics for the coarse states.

We now have completed the construction of all the necessary structures to introduce our effective theories:
\begin{quote}
	\emph{An effective theory $\mathcal T_{n,p}$ is defined by the Hilbert space $\mathcal H_{(n,p)}$, a Hamiltonian operator $\check H_n[\{\lambda_c\}]$ annihilating the vacuum state $\Omega_{\rm AL} \in \mathcal H_{(n,p)}$, and the operator sub-algebra $\mathfrak U_{(n,p)}$ preserving $\mathcal H_{(n,p)}$; and we write $\mathcal T_{n,p} = \big( \mathcal H_{(n,p)}, \check H_n, \Omega_{\rm AL} \big)$.}
\end{quote}
Each effective theory $\mathcal T_{n,p}$ represents the coarse-graining of the fundamental canonical theory at resolution parameters (or scales) $(n,p)$. These scales correspond to the maximal numbers of vertices and edges connecting pairs of vertices respectively in the graphs of the states in $\mathcal H_{(n,p)}$, and we interpret them as ultraviolet cutoffs for the theory.

The final picture is that starting from the fundamental LQG theory, we arrive at a discrete set $\mathcal T$ of effective theories $\mathcal T_{n,p}$, with the parameters $n,\, p \in \mathbbm N^+ \cup \{\infty\}$. This set is naturally endowed with a partial order  "$\prec$" induced by the componentwise partial order on $(\mathbbm N^+ \cup \{\infty\})^2$, i.e.~$(n,p) \prec (m,q)$ if and only if $n \leq m$ and $p \leq q$. This structure enables us to introduce a collection of coarse-grained models for canonical loop quantum gravity, which we define in the next section. Furthermore, we will show how the coarse-graining maps $\mathcal I_{n,p}$ can be used to implement a Hamiltonian non-perturbative renormalization scheme for these coarse-grained models and formulate renormalization flow equations.

\section{Coarse-grained models and Hamiltonian renormalization}\label{sec:Hamiltonian_Renormalization}

Using the notion of effective theories introduced above and their set structure, we define coarse-grained models as follows:
\begin{quote}
	\emph{a coarse-grained model is an ordered family of effective theories $\mathcal T_{n,p}$, i.e.~an ordered subset of the poset $\mathcal T$.}
\end{quote}
An effective theory $\mathcal T_{n,p}$ within a coarse-grained model is understood to be providing a description of a possible continuum theory at fixed scales $(n,p)$. 
The existence of a continuum theory within a coarse-grained model requires the effective theories at different scales to be consistent with each other. In this context, consistency means that when one coarse-grains an effective theory at a resolution $(n,p)$ to a resolution $(m,q)\prec(n,p)$, one obtains the effective theory at scale $(m,q)$. This is where the Hamiltonian non-perturbative renormalization framework comes in; see \cite{RodriguezZarate:2025ipb, Zarate:2025zqz} and the references therein. The goal of this framework is to impose this consistency through renormalization flow equations for the Hamiltonian dynamics. Their solutions, i.e.~ultraviolet fixed points, provide the values of the coupling parameters which guarantee the consistency among the effective theories as well as the existence of a continuum theory.

Our objective is to implement the Hamiltonian non-perturbative renormalization framework, developed and studied in \cite{Lang:2017beo, Lang:2017yxi, Lang:2017oed, Lang:2017xrb, Liegener:2020dbc, Thiemann:2020cuq, Thiemann:2022voq, Thiemann:2022ulb, Thiemann:2022ykh, Zarate:2025qlg, RodriguezZarate:2025ipb, Zarate:2025zqz}, for our coarse-grained models as per the definition above.
This framework mainly addresses two problems: the quantization ambiguities present in the construction of the quantum Hamiltonian operators, which are primarily due to the highly non-polynomial dependence of the classical Hamiltonian functional on the basic fields; and the derivation of the continuum limit of the fundamental loop quantum theory.
Using methods from constructive quantum field theory (see \cite{Thiemann:2020cuq} and references therein), the authors proposed a setting to define a renormalization flow directly on Hamiltonian operators and vacuum states. The fixed points of this flow impose the consistency among the effective theories at different scales and thus define a genuine continuum quantum theory.
These results place the Hamiltonian non-perturbative renormalization as a framework for completing the dynamics of LQG and a concrete tool for understanding its continuum limit. They also constitute a foundation for connecting the canonical formulations of LQG and certain covariant formulations such as spin foams and group field theory.

In short, the ingredients required for the implementation of the Hamiltonian non-perturbative renormalization framework are: (i) an ordered family of theories identified by triples $(\mathcal H_k, \hat H_k, \Omega_k)$, with $k$ ranging in some subset of positive integers, where each triple consists of a Hilbert space $\mathcal H_k$, a Hamiltonian $\hat H_k$ and a vacuum state $\Omega_k$ annihilated by the Hamiltonian; (ii) a set of linear isometric embeddings $\mathcal E_{k,k'}$ with $k<k'$, which map a space $\mathcal H_k$ into $\mathcal H_{k'}$ for every pair of labels satisfying $k<k'$. Additionally, the embedding maps must satisfy the transitivity property: $\mathcal E_{k',k^{\prime\prime}} \circ \mathcal E_{k,k'} = \mathcal E_{k,k^{\prime\prime}}$ for any $k<k'<k^{\prime\prime}$. 
Given these structures, one defines the renormalization flow on the family of theories via the following sequence of equations, labeled by $r\in \mathbbm{N}^+$, 
\begin{align}
	\langle \Omega_k^{(r+1)} , \Psi \rangle_{\mathcal H_k^{(r+1)}} &= \langle \Omega_{m(k)}^{(r)} , \mathcal E_{k,m(k)} \Psi \rangle_{\mathcal H_{m(k)}^{(r)}} \label{General_Vacuum_Renormalization}\\
	\langle \Phi , \hat H_k^{(r+1)} \Psi \rangle_{\mathcal H_k^{(r+1)}} &= \langle \mathcal E_{k,m(k)} \Phi , \hat H_{m(k)}^{(r)} \mathcal E_{k,m(k)} \Psi \rangle_{\mathcal H_{m(k)}^{(r)}} \ , \label{General_Hamiltonian_Renormalization}
\end{align}
for all states $\Psi$ and $\Phi$ in $\mathcal H_k$, where $\langle \cdot \rangle_{\mathcal H_k}$ denotes the scalar products on $\mathcal H_k$ and $m(k)$ is a fixed map on the set of labels of the effective theories satisfying $m(k) > k$. The first equation \eqref{General_Vacuum_Renormalization} identifies the vacuum states when moving from \emph{finer to coarser resolutions}, while the second equation \eqref{General_Hamiltonian_Renormalization} defines the matrix elements of the Hamiltonian operator (or quadratic form) in the coarser theory in terms of the restriction of the matrix elements of the Hamiltonian operator in the finer theory. If a fixed point of the flow equations exists, then it provides a resolution $n_o$ and a fixed set of values for the flow parameters such that the equations \eqref{General_Vacuum_Renormalization} and \eqref{General_Hamiltonian_Renormalization} are satisfied for all effective theories with a resolution finer than $n_o$, and a continuum limit of the model is thereby obtained.

In our context, a family of effective theories is given by a subset of the poset $\mathcal T$ of all the effective theories. Choosing a monotonic function $m(k)$ selects different ordered families of effective theories across $\mathcal T$, and it is equivalent to choosing a coarse-grained model per our definition above. For instance, one could take the function $m$ to be $m(n,p) = (n+1,p)$, which keeps the parameter $p$ fixed. This would define a specific family of coarse-grained models labeled by the value of $p$ (maximal number of edges connecting any two vertices), while the flow is defined with respect to the number of vertices $n$.

The last ingredient for successfully implementing the renormalization framework for our coarse-grained models is the set of linear isometric embeddings $\mathcal E_{(n,p),(m,q)}$, which we call \emph{refinement maps} from now on, that we need to construct using the coarse-graining maps $\mathcal I_{n,p}$ defined in \eqref{Coarse-graining_maps}. In order to achieve this, we begin by analyzing the properties of the coarse-graining maps $\mathcal I_{n,p}$ and their adjoint maps, then we establish a method to define the refinement maps for any model.

\subsection{Construction of the refinement maps for the coarse-grained models}

Consider a coarse-grained model with an ordered family of effective theories $\mathcal T_{n,p}$. For clarity, we map the sequence $\{(n,p)\}$ defining the model to $N^+ \cup \{\infty\}$, and we replace the label $(n,p)$ by $k \in \mathbbm N^+ \cup \{\infty\}$ in the notation of the various objects used in the remainder of this section. The reader should keep in mind that everything that follows is applicable to any coarse-grained model, as per our definition.

We begin with an overview of the properties of the coarse-graining maps defined in \eqref{Coarse-graining_maps}. A map \mbox{$\mathcal I_k$} is closed and bounded with domain $\mathcal D[\mathcal I_k] = \mathcal H_{\rm T}$, and it is a projection onto its image ${\rm Ran}[\mathcal I_k] = \mathcal H_k \subset \mathcal H_{\rm T}$. Hence, the adjoint map $\mathcal I_k^*$ is also a closed and bounded operator with domain $\mathcal D[\mathcal I_k^*] = \mathcal H_{\rm T}$; its kernel is given by ${\rm Ker}[\mathcal I_k^*] = {\rm Ran}[\mathcal I_k]^\perp = \mathcal H_k^\perp$ and therefore it is injective on $\mathcal H_k$. Furthermore, $\mathcal I_k^*$ is a projection onto its image given by ${\rm Ran}[\mathcal I_k^*] = {\rm Ker}[\mathcal I_k]^\perp$.

The first idea to construct an embedding map from a space $\mathcal H_k$ onto a space $\mathcal H_m$ with $k<m$ would be to define the refinement map as $\mathcal E_{k,m}\equiv \mathcal I_m \mathcal I_k^*$. Unfortunately, this does not work for two reasons: first, the maps $\mathcal I_m \mathcal I_k^*$ fail to be isometric embeddings because the maps $\mathcal I_k^*$ are not partial isometries; second, they do not satisfy the transitivity property because it is not clear that the spaces ${\rm Ran}[\mathcal I_k^*]$ satisfy the nesting property ${\rm Ran}[\mathcal I_k^*] \subset {\rm Ran}[\mathcal I_{k+1}^*]$ for any $k\in \mathbbm N^+$. The issue of isometry can be solved easily by taking the polar decomposition of $\mathcal I_k^*$, then using the induced partial isometries and their adjoint maps, instead of $\mathcal I_k$ and $\mathcal I_k^*$, to define the embeddings. However, the nesting property is either true by construction within a given model or it is not, and it is equivalent to the statement that ${\rm Ker}[\mathcal I_{k+1}] \subset {\rm Ker}[\mathcal I_k]$ for any $k\in \mathbbm N^+$.
At present, we can neither prove nor disprove this statement for a generic coarse-grained model. Consequently, we use a less straightforward route to arrive at proper refinement maps, which we develop below.

For each space $\tilde{\mathcal H}_\Gamma \subset \mathcal H_{\rm T}$, we identify the \emph{smallest} (in the sense of the inclusion relation) coarse Hilbert space $\mathcal H_{\sigma_\Gamma}$, in the family defining the model under consideration, which contains $\tilde{\mathcal H}_\Gamma$. Then we define a family of maps labeled by $k \in \mathbbm N^+$ as
\begin{align}
	\mathcal C_\Gamma^{(k)} &: \tilde{\mathcal H}_\Gamma \longrightarrow \mathcal H_k \notag \\[1em]
	\mathcal C_\Gamma^{(k)} &\equiv 
	\begin{cases}
		\prod\limits_{i=k}^{\underrightarrow{\sigma_\Gamma}} \mathcal I_i = \mathcal I_k \dots \mathcal I_{\sigma_\Gamma}, \qquad &\text{if }k\leq\sigma_\Gamma \\[1em]
		\prod\limits_{i=\sigma_\Gamma}^{\underrightarrow{k}} \mathcal I_i = \mathcal I_{\sigma_\Gamma} \dots \mathcal I_k = \mathcal I_{\sigma_\Gamma}, \qquad &\text{if }k\geq\sigma_\Gamma
	\end{cases}\ ,
	\label{C_maps}
\end{align}
where the last equality in the second line of \eqref{C_maps} follows from the given order in the model and the property \eqref{Hnp_nesting} of the coarse-graining maps $\mathcal I_k$.
From the definition \eqref{C_maps} we obtain that the maps $\mathcal C_\Gamma^{(k)}$ satisfy
\begin{align}
	\mathcal C_\Gamma^{(k)} = \mathcal I_k \mathcal C_\Gamma^{(k+1)}, \qquad \text{for any }k \in \mathbbm N^+ \ .
\end{align}

We hence define a family of maps on $\mathcal H_{\rm T}$ as
\begin{align}
	\mathcal C_k \equiv \underset{\Gamma}{\otimes}\, \mathcal C_{\Gamma}^{(k)} \ .
	\label{Enhanced_Coarse-graining_maps}
\end{align}
We can then see that, unlike the maps $\mathcal I_k$, the maps $\mathcal C_k$ (short for $\mathcal C_{n,p}$) apply the coarse-graining procedure in steps. Namely, the resolution scale is decreased iteratively until the target scale is reached. In this regard, although it may seem less efficient, an iterative coarse-graining process could present computational advantages when calculating the image of a state.

It follows from the properties of the maps $\mathcal I_k$ that the maps $\mathcal C_k$ are closed and bounded with domain $\mathcal D[\mathcal C_k] = \mathcal H_{\rm T}$, and they also satisfy
\begin{align}
	\mathcal C_k = \mathcal I_k \mathcal C_{k+1}, \qquad \text{for any }k \in \mathbbm N^+ \ .
\end{align}
This directly implies that
\begin{align}
	{\rm Ker}\qty[\mathcal C_m] \subset {\rm Ker}\qty[\mathcal C_k], \qquad \text{for any }m>k \ ,
	\label{Nesting_property}
\end{align}
which is exactly the property that we were missing for the maps $\mathcal I_k$.
Indeed, thanks to \eqref{Nesting_property}, the closures $\mathcal L_k$ of the images of the adjoint maps $\mathcal C_k^*$, which are also closed and bounded maps with domain $\mathcal D[\mathcal C_k^*] = \mathcal H_{\rm T}$, satisfy the nesting property: $\mathcal L_k \subset \mathcal L_{k+1}$.
However, the adjoint maps $\mathcal C_k^*$ are not partial isometries. Nevertheless, as we suggested earlier, we can use their polar decompositions to obtain partial isometries which we employ to define the refinement maps. 

The polar decomposition theorem implies that for every map $\mathcal C_k^*$ there exists a unique partial isometry $\mathcal U_k$, equivalently $\mathcal U_k^*$ for the sake of clearer notation, such that
\begin{align}
	\mathcal C_k^* = \mathcal U_k^* \sqrt{\mathcal C_k \mathcal C_k^*}\ ,
\end{align}
with initial space $\mathcal H_k$, ${\rm Ker}[\mathcal U_k^*] = {\rm Ker}[\mathcal C_k^*]$ and ${\rm Ran}[\mathcal U_k^*] = {\rm Ran}[\mathcal C_k^*] = \mathcal L_k$. Furthermore, the adjoint map $\mathcal U_k$ is a partial isometry with initial space $\mathcal L_k$, and we have ${\rm Ker}[\mathcal U_k] = {\rm Ker}[\mathcal C_k]$ and ${\rm Ran}[\mathcal U_k] = \mathcal H_k$. From these properties it follows that:
\begin{itemize}
		
	\item $\mathcal U_k \mathcal U_k^*$ is an orthogonal projection onto $\mathcal H_k$ and $(\mathcal U_k \mathcal U_k^*)|_{\mathcal H_k} = \mathbbm 1_{\mathcal H_k}$;
	
	\item $\mathcal U_k^* \mathcal U_k$ is an orthogonal projection onto $\mathcal L_k$ and $(\mathcal U_k^* \mathcal U_k)|_{\mathcal L_k} = \mathbbm 1_{\mathcal L_k}$.
\end{itemize}
The above statements hold for any $k\in \mathbbm N^+$; hence, we can define a family of embedding maps as
\begin{align}
	\mathcal E_{k,m}: \mathcal H_k \longrightarrow \mathcal H_m, \qquad \mathcal E_{k,m}\equiv \mathcal U_m \mathcal U_k^*, \qquad \text{for any }m>k \in \mathbbm N^+ \ .
	\label{Refinement_maps}
\end{align}
These maps are isometric thanks to the fact that $\mathcal U_m$ and $\mathcal U_k^*$ are partial isometries, and we can show that the maps $\mathcal E_{k,m}$ satisfy the transitivity property: given any positive integers such that $k_1<k_2<k_3$ we have
\begin{align}
	\mathcal E_{k_2,k_3} \mathcal E_{k_1,k_2} = \mathcal U_{k_3} \mathcal U_{k_2}^* \mathcal U_{k_2} \mathcal U_{k_1}^* = \mathcal U_{k_3} \mathcal U_{k_1}^* = \mathcal E_{k_1,k_3}\ ,
\end{align}
where the second equality follows from the fact that $\mathcal U_{k_2}^* \mathcal U_{k_2}$ is an orthogonal projection onto $\mathcal L_{k_2}$, combined with the nesting property which guarantees that ${\rm Ran}[\mathcal U_{k_1}^*] = \mathcal L_{k_1} \subset \mathcal L_{k_2}$.

We therefore have shown that the maps $\mathcal E_{k,m}$ defined in \eqref{Refinement_maps} satisfy all the necessary requirements for proper refinement maps between the coarse Hilbert spaces $\mathcal H_k$. Hence, we take the embedding maps $\mathcal E_{k,m}$ to be the refinement maps for the definition of the Hamiltonian renormalization flow equations in our coarse-grained models.

\subsection{Renormalization flow equations for the coarse-grained models}

Now that we have appropriate refinement maps $\mathcal E_{k,m}$, which isometrically embed the coarse Hilbert spaces $\mathcal H_k$ into each other, we can finally introduce the Hamiltonian renormalization flow equations, corresponding to \eqref{General_Vacuum_Renormalization} and \eqref{General_Hamiltonian_Renormalization}, for the coarse-grained model under consideration. 
Namely 
\begin{align}
	\langle \Omega_{\rm AL}^{(r+1)} , \Psi \rangle_{\mathcal H_k^{(r+1)}} &= \langle \Omega_{\rm AL}^{(r)} , \mathcal E_{k,k+1} \Psi \rangle_{\mathcal H_{k+1}^{(r)}} \label{Vacuum_Renormalization_eq}\\
	\langle \Phi , \check H_k^{(r+1)}[\{\lambda_c\}]\, \Psi \rangle_{\mathcal H_k^{(r+1)}} &= \langle \mathcal E_{k,k+1} \Phi , \check H_{k+1}^{(r)}[\{\lambda_c\}]\, \mathcal E_{k,k+1} \Psi \rangle_{\mathcal H_{k+1}^{(r)}}\ . \label{Hamiltonian_Renormalization_eq}
\end{align}
These equations dictate the flow between the effective theories $\big( \mathcal H_k, \check{H}_k, \Omega_{\rm AL} \big)$. The important observation is that the vacuum renormalization equations are trivially satisfied in our case, since by construction the coarse-graining maps $\mathcal I_k$ map the vacuum state $\Omega_{\rm AL}$ to itself and consequently for any $m>k>0$: $\mathcal E_{k,m}\Omega_{\rm AL} = \mathcal E_{k,m}^*\Omega_{\rm AL} = \Omega_{\rm AL}$. Hence, one would only need to solve the equations for the matrix elements of the effective Hamiltonians $\check H_k$ defined in \eqref{Effective_Hamiltonian}. These equations are to be solved for the coupling coefficients $\{\lambda\}$ in \eqref{Effective_Hamiltonian} as functions of the scaling parameter(s) $k =(n,p)$. 
Note that in the definition of an effective Hamiltonian \eqref{Effective_Hamiltonian}, we have $(n+7)$ coupling coefficients. This means that the number of these parameters grows linearly with the cutoff $n$, which can go to infinity. Nevertheless, when considering a specific coarse-grained model, one must reduce the number of independent couplings to a fixed and finite value $\nu_c$ in order to make sense of the renormalization flow equations. Hence, the coarse-grained model must be further characterized by the selected independent couplings, together with the expressions of the remaining couplings in terms of these.

The complexity of the refinement maps makes solving equations \eqref{Hamiltonian_Renormalization_eq} analytically challenging; nonetheless, we have shown that a renormalization flow can be defined for a generic coarse-grained model satisfying our definition. This completes the implementation of the Hamiltonian non-perturbative renormalization framework for all coarse-grained models built from the effective theories $\mathcal T_{n,p} = \big( \mathcal H_{(n,p)}, \check H_n, \Omega_{\rm AL} \big)$.


\section{Summary \& outlook}

In this article, we presented the construction of a family of effective theories and the induced coarse-grained models for canonical loop quantum gravity, with an explicit implementation of the Hamiltonian non-perturbative renormalization framework for these models.
Our construction rests on a coarse-graining procedure for spin network states which preserves the number of independent loops. This procedure uses a new notion of abstract graphs with geometric tags and it extends the vertex coarse-graining method via gauge-fixing \cite{Livine:2013gna, Charles:2016xwc} with two additional methods developed here: a systematic prescription for coarse-graining multiple edges connecting any pair of vertices down to a fixed number $p$ of coarse edges, and a method for coarse-graining the geometric tags which we introduced. These four operations are put together to build coarse-graining maps $\mathcal I_{n,p}$ that send kinematical states of LQG to states supported on multi-reflexive \mbox{$(n,p)$-complete} graphs -- graphs with exactly $n$ vertices and in which every pair of vertices is connected by exactly $p$ edges, in addition to an arbitrary number of self-loops. For each choice of $n,\, p\in \mathbbm N^+ \cup \{\infty\}$, this construction yields a separable coarse Hilbert space $\mathcal H_{(n,p)}$, and the parameters $(n,p)$ are interpreted as resolution scales and ultraviolet cutoff.

Furthermore, we proceed to define an effective dynamics through the identification and classification of the effective interactions induced by fundamental Hamiltonian operators of LQG on the coarse states. By examining the interplay between the actions of the various operators and our coarse-graining procedure, we concluded that each operator admits a well-defined effective counterpart on the coarse Hilbert spaces, and that these effective operators organize naturally into a hierarchy of vertex-interactions, distinguished further by whether they preserve or change the graph of the coarse state. This classification provides a systematic characterization of the interaction terms available to obtain an effective Hamiltonian for LQG. On this basis, we proposed a general effective Hamiltonian $\check H_n$ on each space $\mathcal H_{(n,p)}$, given as a sum over all admissible interactions with coupling coefficients $\{\lambda_c\}$. These coefficients eventually constitute the parameters of the renormalization flow. These ingredients together define an effective theory $\mathcal T_{n,p} \equiv \big( \mathcal H_{(n,p)}, \check H_n, \Omega_{\rm AL} \big)$, and we obtain a poset $\mathcal T$ of effective theories parametrized by the scales $n,\, p\in \mathbbm N^+ \cup \{\infty\}$.

Thanks to these structures, we introduced the coarse-grained models as ordered families of effective theories $\mathcal T_{(n,p)}$, and we developed a proposal to implement the Hamiltonian renormalization framework \cite{Lang:2017beo, Lang:2017yxi, Lang:2017oed, Lang:2017xrb, Liegener:2020dbc, Thiemann:2020cuq, Thiemann:2022voq, Thiemann:2022ulb, Thiemann:2022ykh, Zarate:2025qlg, RodriguezZarate:2025ipb, Zarate:2025zqz} for them. This is achieved using the coarse-graining maps $\mathcal I_{n,p}$, which provide a collection of refinement maps $\mathcal E_{(n,p),(m,q)}$, that satisfy all the properties required to write down the renormalization flow equations for the coupling coefficients of the effective Hamiltonian. Taken together, these results provide a concrete setting in which the Hamiltonian non-perturbative renormalization framework for canonical LQG can be studied explicitly, for a large collection of coarse-grained models. 
Detailed analytic or numerical studies of the renormalization flow equations and their possible solutions are open directions for future work. 
An intermediate step for testing our ideas would be to consider simplified settings, such as the $U(1)^3$ Euclidean quantum gravity \cite{RodriguezZarate:2025ipb}, where the theory is exactly solvable and the calculations for the flow would be more manageable.
All these structures and frameworks can be straightforwardly applied to the Yang-Mills sector coupled to gravity, and it would be interesting to extend the methods developed here to the scalar and fermionic sectors.

Finally, it is worth mentioning that the structure of these effective theories suggests several avenues for connecting to other research directions. In particular, given the multi-reflexive \mbox{$(n,p)$-complete} graph structure, the use of complexifier coherent states \cite{Thiemann:2000bw, Thiemann:2000ca, Thiemann:2000bx, Thiemann:2000by, Thiemann:2002vj, Sahlmann:2001nv} and graph coherent states \cite{Assanioussi:2018zit, Assanioussi:2020fsz} can pave the way for studying semi-classical regimes within an effective theory, and how it could translate into the renormalization flow. Additionally, these effective theories encode the structure of the two-vertex models or \emph{candy graph} models, developed and studied earlier \cite{Borja:2010gn, Aranguren:2022nzn, Cendal:2024uzu, Garay:2025cis, Garay:2025bqk, Assanioussi:2026cee}. We expect that the classical as well as the quantum analyses of these models may provide complementary understanding of the dynamics in our coarse-grained models and their renormalization.


\section*{Acknowledgment}
The authors would like to thank Melissa Rodriguez Zarate for helpful and insightful discussions on the framework of Hamiltonian renormalization.

This research was funded by the National Science Centre, Poland, through grant no.~2023\slash 51\slash D\slash ST2\slash 00296. For the purpose of Open Access, the authors have applied a CC-BY public copyright licence to any Author Accepted Manuscript (AAM) version arising from this submission.


\appendix
\section{Expressions of the quantum operators involved in the Hamiltonian operators}\label{Expressions_Hamiltonian}

We provide the expressions of the operators used in defining the several quantum Hamiltonian operators present in the LQG literature, which in turn are used to define the effective dynamics for the coarse-grained models. As discussed in section \ref{sec:Effective Dynamics}, the effective dynamics is considered within a deparametrized setting; therefore, all the operators must be at least symmetric on suitable domains. When necessary, we consider simple symmetrizations of the operators resulting from the various regularizations, such that the effective Hamiltonians $\check H_n$ in \eqref{Effective_Hamiltonian} are at least densely defined on the Hilbert spaces $\mathcal H_{(n,p)}$.

Given a state in a Hilbert space $\mathcal H_\Gamma$, the sum over $v$ denotes the sum over the vertices in $\Gamma$, while the sum over $I,J,\dots$ denotes the sum over the edges meeting at a given vertex. To simplify the notation we use $J^{I}_i \equiv J_{v,e_I,i}$. We then have the following expressions:
\begin{itemize}
	\item Volume operator $\hat V$, for which there are two versions:
	
	- The Ashtekar-Lewandowski volume operator introduced in \cite{Ashtekar:1997fb}
	\begin{equation}\label{V_AL}
		\hat{V}_{\rm AL}[N]\equiv(\kappa\beta)^{3/2}\sum_{v}N(v)\sqrt{\frac{1}{48}\abs{\sum_{I,J,K}\epsilon(e_I,e_J,e_K)\varepsilon^{ijk}J^{I}_i J^{J}_j J^{K}_k}}\ ,
	\end{equation}
	where the factor $\epsilon(e_I,e_J,e_K)$ stands for the determinant of the tangent vectors to the edges $e_I$, $e_J$ and $e_K$ in this order.
	
	- The Rovelli-Smolin volume operator introduced in \cite{Rovelli:1994ge}
	\begin{equation}\label{V_RS}
		\hat{V}_{\rm RS}[N]\equiv(\kappa\beta)^{3/2}\sum_{v}N(v)\sqrt{\frac{1}{48}\sum_{I,J,K}\abs{\varepsilon^{ijk}J^{I}_i J^{J}_j J^{K}_k}}\ .
	\end{equation}
	
	\item Ricci scalar operator $\hat{H}_R$ introduced in \cite{Alesci:2014aza}:
	
	\begin{equation}\label{H_R}
		\hat{H}_R[N]\equiv (\kappa\beta)^2 \sum_{v}N(v)\kappa_R(v) \sum_{I,J}\frac{1}{2}\sqrt{\mathcal{V}^{-1}_v}\sqrt{\delta^{ij}Y^{IJ}_i Y^{IJ}_j}\qty(\frac{2\pi}{c_{IJ}}-\pi+\theta^{IJ}_{v})\sqrt{\mathcal{V}^{-1}_v}\ ,
	\end{equation}
	where $\kappa_R(v)$ is an averaging coefficient, $c_{IJ}$ is a fixed parameter, $\mathcal{V}^{-1}_v \equiv \lim_{\epsilon\rightarrow0}(\hat{V}^2_v+\epsilon\kappa^{3/2})^{-1}\hat{V}_v$ with $\hat{V} = \hat{V}_{\rm AL}$ or $\hat{V}_{\rm RS}$, and
	\begin{equation}
		\theta^{IJ}_{v} \equiv \cos^{-1}\qty(\frac{\delta^{ij} J^{I}_i J^{J}_j}{\abs{J^{I}}\abs{J^{J}}}) \,,\quad \abs{J^{I}}\equiv\sqrt{\delta^{kl} J^{I}_k J^{I}_l}\ .
	\end{equation}
	
	\item The Euclidean operator $\hat{H}_E$:
	
	\begin{equation}\label{H_E}
		\hat{H}_E[N]\equiv \sum_{v} N(v) \sum_{I,J} \hat{H}_E[\alpha_{IJ}]\ ,
	\end{equation}
	such that
	\begin{align}\label{H_E[IJ]}
		\hat{H}_E[\alpha_{IJ}] &\equiv \hat{C}_E[\alpha_{IJ}] + \hat{C}_E^*[\alpha_{IJ}]
	\end{align}
	and
	\begin{align}\label{C_E[IJ]}
		\hat{C}_E[\alpha_{IJ}] &\equiv \frac{\mathrm{i}}{\kappa\beta}\frac{1}{W^2_l} \frac{8}{E(v)}\sum_{K} \epsilon(e_I,e_J,e_K) \operatorname{Tr}\qty(D^{(l)}\qty(h_{\alpha_{IJ}}) D^{(l)}\qty(h^{-1}_{s_K})\qty[D^{(l)}\qty(h_{s_K}),\hat{V}])\ ,
	\end{align}
	where $*$ denotes the adjoint operator, $W_l \equiv \sqrt{l(l+1)(2l+1)}$, $E(v)=\binom{w_v}{3}$ with $w_v$ being the valence of $v$, and $s_k$ denotes a segment of the edge $e_K$ starting at $v$. The details of the various regularization procedures can be found in \cite{Thiemann:1996aw, Thiemann:2003zv, Giesel:2006uj, Alesci:2015wla, Assanioussi:2015gka}.
	
	\item The Lorentzian operator $\hat{H}_L$ \cite{Thiemann:1996aw}:
	
	\begin{align}\label{H_L}
		\hat{H}_L[N]\equiv \sum_{v}N(v) \qty(\hat{H}_L[v] + \hat{H}_L^*[v])\ ,
	\end{align}
	such that 
	\begin{align}
		\hat{H}_L[v] \equiv \frac{2\mathrm{i}}{\kappa^5\beta^7}\frac{1}{W^2_l} \frac{8}{E(v)}\sum_{I,J,K}\epsilon(e_I,e_J,e_K)
		\operatorname{Tr}\bigg(
		& D^{(l)}\qty(h^{-1}_{s_I})\qty[D^{(l)}\qty(h_{s_I}),\qty[H_E,\hat{V}_v]] \notag \\
		\times & D^{(l)}\qty(h^{-1}_{s_J})\qty[D^{(l)}\qty(h_{s_J}),\qty[H_E,\hat{V}_v]] \notag \\
		\times & D^{(l)}\qty(h^{-1}_{s_K})\qty[D^{(l)}\qty(h_{s_K}),\hat{V}_v]\bigg)\ .
	\end{align}
	
\end{itemize}


\newpage
\bibliographystyle{apsrev4-2-modified}
\bibliography{main_references.bib}


\end{document}